\documentstyle[a4,11pt]{article}

\newcommand{\sect}[1]{\setcounter{equation}{0}\section{#1}}
\newcommand{\subsect}[1]{\subsection{#1}}
\renewcommand{\theequation}{\thesection.\arabic{equation}}
\newcommand{\vs}[1]{\rule[ #1 mm]{0mm}{#1 mm}}
\newcommand{\hs}[1]{\hspace{#1 mm}}
\newcommand{\lbl}[1]{\label{eq:#1}}
\newcommand{\rf}[1]{(\ref{eq:#1})}
\newcommand{\nn}{\nonumber}
\newcommand{\be}{\vs{2}\begin{eqnarray}}
\newcommand{\ee}{\vs{2}\end{eqnarray}}

\newcommand{\bea}{\large\begin{eqnarray}}
\newcommand{\ena}{\end{eqnarray}\normalsize}
\newcommand{\nnbea}{\large\begin{eqnarray*}}
\newcommand{\nnena}{\end{eqnarray*}\normalsize}
\newcommand{\leqa}{\lefteqn}

\def\ie{{\it i.e.\ }}

\newcommand{\grt}[1]{\mbox{\LARGE #1}}
\newcommand{\lp}{\left(}
\newcommand{\blp}{\biggl(}
\newcommand{\rp}{\right)}
\newcommand{\brp}{\biggr)}
\newcommand{\tld}[1]{\widetilde{#1}}
\newcommand{\ovl}[1]{\overline{#1}}
\newcommand{\lra}{\ \longrightarrow\ }
\newcommand{\Llra}{\ \Longleftrightarrow\ }
\newcommand{\Lra}{\ \Longrightarrow\ }
\newcommand{\mbf}[1]{\mbox{\boldmath $#1$}}
\newcommand{\tens}{\! \otimes \!}
\newcommand{\pt}{\! \cdot \!}

\newcommand{\shalf}{\textstyle{\frac{1}{2}}\displaystyle}
\newcommand{\dps}{\displaystyle}
\newcommand{\nms}{\normalsize}
\newcommand{\sst}{\scriptstyle}
\newcommand{\cm}{\hspace{1cm}}

\newcommand{\tr}{\mbox{tr}}
\newcommand{\fprt}{\mbox{\boldmath $\partial$}}
\newcommand{\fbprt}{\ovl{\mbox{\boldmath $\partial$}}}
\newcommand{\fc}{\mbox{\boldmath $c$}}
\newcommand{\bc}{\ovl{c}}
\newcommand{\bz}{\ovl{z}}
\newcommand{\bC}{\ovl{C}}
\newcommand{\bZ}{\ovl{Z}}
\newcommand{\zbz}{(z,\ovl{z})}
\newcommand{\ZBZ}{(Z,\ovl{Z})}
\newcommand{\ZBZp}{(Z',\ovl{Z}')}
\newcommand{\prt}{\partial}
\newcommand{\bprt}{\ovl{\partial}}
\newcommand{\paz}{\partial_z}
\newcommand{\paZ}{\partial_Z}
\newcommand{\pabz}{\partial_{\ovl{z}}}
\newcommand{\pabZ}{\partial_{\ovl{Z}}}
\newcommand{\hdelta}{\widehat{\delta}\,}
\newcommand{\DER}[1]{\frac{\prt}{\prt #1 \zbz}}
\newcommand{\DERR}[2]{\frac{\prt #1 \zbz}{\prt #2 \zbz}}

\newcommand{\mub}{\ovl{\mu}}
\newcommand{\lam}{\lambda}
\newcommand{\lamb}{\ovl{\lambda}}
\newcommand{\Lam}{\Lambda}
\newcommand{\al}{{\alpha}}

\newcommand{\jb}{{\ovl{\jmath}}}
\newcommand{\ib}{{\ovl{\imath}}}
\newcommand{\lb}{{\ovl{l}}}

\newcommand{\rb}{{\ovl{r}}}

\newcommand{\cL}{{\cal L}}
\newcommand{\TH}{\mbox{\Large ${\dps \vartheta}$}}
\begin{document}

\setcounter{page}{0}
\renewcommand{\thefootnote}{\fnsymbol{footnote}}
\pagestyle{empty}

\begin{center}

{\LARGE {\bf Kodaira-Spencer deformation of complex structures}}\\[3mm]
{\LARGE {\bf and Lagrangian field theory}}\\[1cm]

{\large Giuseppe BANDELLONI $^a$
\footnote{e-mail : {\tt beppe@genova.infn.it}}
and Serge LAZZARINI $^b$
\footnote{e-mail : {\tt sel@cpt.univ-mrs.fr}}
}\\[7mm]
 
{\em $^a$ Istituto di Fisica della Facolt\`a di Ingegneria, Universit\`a degli 
Studi di Genova,\\ P.le Kennedy, I-16129 GENOVA, Italy\\ and\\
Istituto Nazionale di Fisica Nucleare, INFN, Sezione di Genova,\\
via Dodecaneso 33, I-16146 GENOVA, Italy}\\
 
\vskip 1.cm
 
{\em $^b$ Centre de Physique Th\'eorique, CNRS-Luminy,\\
Case 907, F-13288 MARSEILLE Cedex 9, France.}
\end{center}
 
\indent
 
\indent
 
\centerline{\large {\bf Abstract}}
 
In complete analogy with the Beltrami parametrization of 
complex structures on a (compact) Riemann surface, we use in this 
paper the Kodaira-Spencer deformation
theory of complex structures on a (compact) complex manifold of higher
dimension. According to the Newlander-Nirenberg theorem, a smooth
change of local complex coordinates can be implemented with respect to an 
integrable complex structure parametrized by a Beltrami differential. 
The question of constructing a local field theory on a
complex compact manifold is addressed and the action of smooth
diffeomorphisms is studied in the BRS algebraic approach. The BRS
cohomology for the diffeomorphisms gives generalized 
Gel'fand-Fuchs cocycles provided that the Kodaira-Spencer
integrability condition is satisfied. The
diffeomorphism anomaly is computed and turns out to be holomorphically
split as in the bidimensional Lagrangian conformal models. Moreover, its
algebraic structure is much more complicated than the one proposed in 
a recent paper \cite{LMNS97}.


\vfill

\noindent{1995PACS classification : 11.10.Gh,03.70}

\noindent{CPT-98/P.3622}

\newpage
 
\renewcommand{\thefootnote}{\arabic{footnote}}
\setcounter{footnote}{0}
\pagestyle{plain}


\sect{Introduction}

\indent

A class of bidimensional conformal models, thanks to the equivalence
between conformal and complex structures on a 2-d manifold,
has been suitably incorporated in the framework of the local field theory 
upon using the Beltrami parametrisation of complex
structures. Moreover, in this parametrisation the holomorphic
factorization of the partition functions of these Lagrangian conformal
models (free bosonic string, $bc$ system possibly coupled to
Yang-Mills gauge fields) has been proved \cite{BB87,Bec88,KLT90,KLS91a,KLS91b}.

In real even dimensions larger than two, the equivalence between conformal
and complex structures is no longer valid. However, complex structures
on a complex manifold are still parametrized by a vector-valued
$(0,1)$-form fulfilling both an integrability and algebraic
conditions. This is the Kodaira-Spencer deformation theory of complex
structures on a manifold which generalizes the Beltrami
parametrization to larger dimensions. 
While in two-(real) dimensions, the Beltrami differential 
has been used in the study of some bidimensional Lagrangian 
conformal models on a (compact) Riemann surface, the Kodaira-Spencer
deformation of complex structures has occured in theoretical 
physics in the context of mirror symmetry in string field models with 
values in some target K\"{a}hler manifold \cite{Vaf94,BCOV94,BS94,LM94}.

Quite recently, the holomorphicity of the partition function for
chiral lagrangians of higher dimensional $bc$ systems on a K\"ahler
manifold has been considered in \cite{LMNS97}, but
only with respect to the so-called ``chiral diffeomorphisms''.
This restriction does not include the whole symmetry 
under any arbitrary reparametrization.
Moreover, this {\em a priori} holomorphicity requirement in the Beltrami
differential defining an (integrable) complex structure is put by
hand, and it is not straightforward to decide whether a higher dimensional 
analogue of the $bc$ system actually enjoys the holomorphic
factorization property. Finally, the proposed ``chiral anomaly''
in \cite{LMNS97} turns out to be much more complicated as it will 
be shown in the paper.

In the present work, we simply address the question whether this 
parametrization might give rise to interesting features for possible 
local field theories on a compact complex manifold, 
for instance, whether the Beltrami differential may play a role at the
field theory level. In particular, we have in mind the four-(real) 
dimensional interesting case and possible links with the recent work
\cite{Zuc97}. For other noteworthy implications of holomorphicity in
the Beltrami differential the reader is referred to \cite{LMNS97}.

Following closely the bidimensional approach, the action of 
smooth diffeomorphisms (connected to the identity) is studied in the 
BRS formulation. In doing so, one has 
the advantage of dealing with the full invariance under
reparametrizations instead of the ``chiral
diffeomorphisms'' as treated in \cite{LMNS97}. Furthermore, the BRS
cohomology for diffeomorphisms is analysed along the line developed in 
\cite{Ba88,BaLa93,BaLa95}, and allows one to compute the local 
expression of a diffeomorphism anomaly.
A few possible Lagrangian models will be also investigated.
The first part of the present paper is devoted to some notation and, according
to the book by K.~Kodaira \cite{Kod86}, we will introduce the
parametrisation of complex structures on a compact manifold 
by vector-valued $(0,1)$-forms, the higher dimensional analogues of
the Beltrami differentials. We shall be content into
generalized matter fields considered as usual tensors.

The main part of the paper is based on the study of the diffeomorphism
cohomology and the finding of anomalies under the locality principle. 
The cohomology can be carried out for the BRS operator of the
diffeomorphisms which is nilpotent under the restriction that the
complex structure is integrable.

\indent

\sect{Basic idea for the perturbation of complex structures}

\indent

Consider a complex compact manifold of complex dimension $n$ 
according to a finite complex analytic covering 
$\{(U_\al,z_\al)\}$ with $M= {\displaystyle \cup_\al}U_\al$
and $z_\al:U_\al\lra\mbox{\nms {\bf C}}^{n}$, the local complex coordinate mapping,
$z\equiv (z^1,\dots,z^n) = (z^k)$, with biholomorphic transition
functions
\be
\mbox{\nms in } U_{\alpha} \cap U_{\beta}\neq \emptyset\ :\
z_\al\ =\ f_{\al\beta}(z_{\beta})\ ,
\ee
for any point in the intersection.
When the underlying smooth structure is considered (thus  
$M$ is viewed as a real smooth manifold of dimension $2n$), the complex
conjugate coordinates $\bz\equiv (\bz^1,\dots,\bz^n)=(\bz^k)$ is regarded
as independent with respect to $z$.

Partial derivatives will be shortly denoted as $\prt/\prt z^k\equiv\prt_k$ 
and $\prt/\prt\bz^j\equiv \prt_\jb$.
We shall denote by $T^{1,0}$ the holomorphic tangent bundle of $M$, and
$\prt$ will stand for the natural frame on $T^{1,0}$
associated to the local background complex coordinates $\zbz$.  
Similarly, $dz$ will be the natural dual basis in the holomorphic
cotangent space $(T^{1,0})^*$,
\be
\prt\ \equiv\ \left(
\begin{array}{c} \prt_1\\ \vdots\\ \prt_n \end{array} \right)\ ,\
\bprt\ \equiv\ \left(
\begin{array}{c} \prt_{\ovl{1}}\\ \vdots\\ \prt_{\ovl{n}} 
\end{array} \right)\ ,\cm 
dz\ \equiv\ (dz^1,\dots,dz^n)\ ,\ d\bz\ \equiv\ (d\bz^1,\dots,d\bz^n)\ ,
\ee
while we have for the external derivative (denoted here by a boldface symbol)
\be
d\ =\ \fprt + \fbprt \ =\ dz\pt\prt+d\bz\pt\bprt\ =\
dz^k \prt_k + d\bz^j \prt_\jb, \cm 
d^2 = \fprt^2 = \fbprt^2 = \left\{ \fprt,\fbprt \right\} = 0\ .
\ee
This system of local complex coordinates (which is fixed once
and for all) will serve as
background for the perturbation implemented by modifying
transition functions, according to the Kodaira-Spencer's approach, as
follows. On any overlapping $U_{\alpha} \cap U_{\beta}\neq \emptyset$,
of the finite covering, 
the background transition function $f_{\al\beta}(z_{\beta})$ is
replaced by another function $f_{\al\beta}(z_{\beta},t))$, such that
$f_{\al\beta}(z_{\beta},0)=f_{\al\beta}(z_{\beta})$, and where $t$ is
a complex parameter belonging to some domain of {\bf C}$^m$ containing
the origin.
One thus obtains a complex analytic family parametrized by $t$ 
of complex compact manifolds. It turns out that the infinitesimal 
deformation of $M$ can be represented by an element of the cohomology 
group $H^1(M,\Theta)$ of $M$ with coefficients in the sheaf 
$\Theta$ of germs of holomorphic vector fields on $M$.
In most cases, the Kodaira-Spencer's conjecture holds, so that 
dim$H^1(M,\Theta) = m$ gives the dimension of the moduli
space of $M$ when the number of moduli can be defined.
A question of primary importance is to know whether it is always possible to
deform a complex manifold. A partial answer is given for instance in
\cite{Kod86}~: It is always possible to deform the complex structure on
any complex compact manifold. 
Furthermore, all the complex compact manifolds belonging to the same complex
analytic family are all diffeomorphic since they are subordinated to
the same smooth structure underlying
the background complex structure. This situation allows one to tackle the
deformation problem by parametrizing the complex structures by a 
vector-valued $(0,1)$-form on $M$ 
which extends to the higher dimensional case
the well-known Beltrami parametrization of complex structures on a
Riemann surface. 

\subsect{The Newlander-Nirenberg theorem and 
the Kodaira-Spencer integrability condition}

\indent

Let $\mbf{\mu}$ be a vector-valued (0,1)-form, a smooth section of 
$T^{1,0} \tens (T^{0,1})^*$ locally represented as
\be
\mbf{\mu} \zbz\ =\ d\bz\pt\mu\zbz\pt\prt\ =\
\mu^k_\jb\zbz d\bz^j \tens \prt_k\ , \cm \mu\ =\ (\mu^k_\jb)\ .
\lbl{mu}
\ee
This object generalizes the usual Beltrami differential but will be
subject to an integrability condition as stated in the 

\indent

\noindent
{\bf {Theorem}} (Newlander-Nirenberg \cite{Kod86}). 
For a given smooth vector-valued (0,1)-form, $\mbf{\mu}$, viewed as a
differential operator of first order and locally defined on $M$ by eq.\rf{mu},
consider the following first order partial differential operator,
\be
\mbf{\cal L}\ =\ \fbprt - \mbf{\mu}\ =\ d\bz\pt(\bprt - \mu\pt\prt)
\ =\ d\bz^i\cL_\ib \ =\ d\bz^i(\prt_\ib - \mu^k_\ib\prt_k)\ .
\lbl{L}
\ee
Next suppose that $\mbf{\mu}$ obeys the following two conditions
\be
\det(I - \mu\mub)\neq 0, \cm \mbox{and} \cm
\mbf{\cal L}\mbf{\mu} \ =\ \fbprt \mbf{\mu} - \mbf{\mu}^2\ =\
 \fbprt \mbf{\mu} - \shalf[\mbf{\mu},\mbf{\mu}]\ =\ 0\ ,
\lbl{integ}
\ee
where the graded bracket reflects the bracket of vector fields.

Hence, the differential operators 
$\cL_{\ovl{1}},\dots,\cL_{\ovl{n}},\ovl{\cL}_1,\dots,\ovl{\cL}_n$ are
linearly independent and they fulfill the condition
\be
\mbf{\cal L}^2\ =\ 0 \Llra
\left[ \cL_\ib,\cL_\jb \right]\ =\ 0, \cm \ib,\jb =1,\dots,n. 
\ee
Moreover, the system of $n$ PDE's
\be
\mbf{\cal L}f\ =\ 0 \Llra \cL_\ib\, f \ =\ 0, \cm \ib =1,\dots,n,
\lbl{pde}
\ee                          
has locally $n$ linearly independent smooth solutions
$Z=(Z^1\zbz,\dots,Z^n\zbz)$, \ie the Jacobian never vanishes,
\be
{\cal J}\equiv\frac{\prt(Z,\bZ)}{\prt(z,\bz)}\ =\ 
|\det \lam|^2 \det(I - \mu\mub) \neq 0
\Lra \det \lam \neq 0\ ,
\lbl{Jac}
\ee
where $\lam = (\prt_i Z^k)$ is the (half)-Jacobian matrix.

\subsect{The smooth change of local complex coordinates}

The solution $Z$ provides a new system of complex coordinates on
$M$, that is a new integrable complex structure parametrized by $\mu$.
The new complex structure $\ZBZ$ on $M$ can be seen as a deformation of the
background complex structure $\zbz$, and each such $\mu$ determines an
integrable complex structure on $M$.

\indent

\noindent
{\bf Remark.} The coordinates $Z$, local smooth 
solutions of the system \rf{pde}, are
defined up to an holomorphic function $F(Z)$.
Also, when $\mu \equiv 0$ the change of local coordinates becomes
holomorphic and belongs to the complex structure
given by $z$. Moreover, one has $\det \lam \neq\ 0$.

\indent

The Newlander-Nirenberg theorem infers a {\em smooth} change of local
complex coordinates in {\bf C}$^n$, $\zbz \lra \ZBZ$. 
These complex coordinates $Z$ can be
re-obtained by the following row-vector of 1-forms
\be
d Z\ =\ \fprt Z + \mbf{\mu} Z\ =\ (dz + d\bz\pt\mu)\pt\lam\ ,
\lbl{dZ}
\ee
one accordingly gets by closure the two following integrability conditions, 
one on $\lam=(\prt_i Z^k)$, $\det \lam \neq 0$, 
\be
\fbprt \fprt Z + \fprt(\mbf{\mu}Z)\ =\ 0 \Llra \bprt \lam - \prt
(\mu\pt\lam)\ =\ 0\ ,
\lbl{compa}
\ee
and the other, the necessary integrability condition \rf{integ} on $\mbf{\mu}$.
Therefore, the defining system of $n$ PDE's \rf{pde} for $Z$, namely
\be
\mbf{\cal L}Z\ =\ (\fbprt - \mbf{\mu}) Z\ =\ 0 \ , \cm
\fbox{$\fbprt \mbf{\mu} - \mbf{\mu}^2\ =\ 0$} \ ,
\lbl{beltra}
\ee
is the analogue of the Beltrami equation but with in addition
an integrability condition.
The latter looks like the vanishing of a curvature.
One may also assume that $\det(I - \mu\mub)>0$ and $\det \lam >0$,
when orientation-preserving coordinate transformations are in order.
    
According to definition \rf{dZ} and its complex conjugate,
one can write in matrix notation
\be
\lp dZ,d\bZ \rp \ =\ \lp dz,d\bz \rp 
\lp \begin{array}{c}
I \\[1mm] \hline \\[-5mm] \mu \end{array} \right|
\left. \hs{-1.2}\begin{array}{c}
\mub \\[1mm] \hline \\[-5mm] I \end{array} \rp
\lp \begin{array}{c}
\lam \\[1mm] \hline \\[-5mm]  0 \end{array} \right|
\left. \hs{-1.2}\begin{array}{c}
0 \\[1mm] \hline \\[-5mm] \lamb \end{array} \rp\ ,
\lbl{matdZ}
\ee
so that by duality
\be
\lp \begin{array}{c} \paZ \\ \pabZ \end{array} \rp \ =\
\lp \begin{array}{c}
\lam \\[1mm] \hline \\[-5mm]  0 \end{array} \right|
\left. \hs{-1.2}\begin{array}{c}
0 \\[1mm] \hline \\[-5mm] \lamb \end{array} \rp ^{-1}
\lp \begin{array}{c}
I \\[1mm] \hline \\[-5mm] \mu \end{array} \right|
\left. \hs{-1.2}\begin{array}{c}
\mub \\[1mm] \hline \\[-5mm] I \end{array} \rp^{-1}
\lp \begin{array}{c} \paz \\ \pabz \end{array} \rp \ .
\ee 
One explicitly finds 
\be
\lp \begin{array}{c} \paZ \\ \pabZ \end{array} \rp \ =\
\lp \begin{array}{c} \lam^{-1}\pt (I - \mub\pt\mu) ^{-1} \pt
\lp \paz - \mub \pt\pabz \rp \\
\lamb^{-1} \pt (I - \mu\pt\mub)^{-1}\pt 
\lp \pabz - \mu\pt \paz \rp 
\end{array} \rp \equiv
\lp \begin{array}{c} \lam^{-1}\pt \nabla \\
\lamb^{-1} \pt \ovl{\nabla} \end{array} \rp  \ .
\lbl{nabla}
\ee

\indent

\sect{The diffeomorphism action}

\indent

By computing the pull-back of the 1-form \rf{dZ} under any smooth
diffeomorphism $\varphi$, $\varphi^* d Z = d \varphi^*Z =
(d \varphi^*w + d \varphi^*\ovl{w})\pt(\varphi^*\mu))\pt (\varphi^*\lam)
= (dz + d\bz\pt(\mu^{\varphi}))\pt(\lam^{\varphi})$, where $\zbz$ are
coordinates in the source chart of $\varphi$ and
$(w,\ovl{w})=\varphi\zbz$ in the target chart,
the finite diffeomorphism transformation of the matrices of basic 
geometric objects reads
\bea
\lam^{\varphi} &=& \blp \prt\varphi + 
\prt\ovl{\varphi}\pt(\mu\circ\varphi)\brp \pt (\lam\circ\varphi)
\ ,\nn\\[-4mm]
&&\lbl{finite}\\[-2mm]
\mu^{\varphi} &=& \blp \bprt\varphi + \bprt\ovl{\varphi}\pt(\mu\circ\varphi)
\brp \pt \blp \prt\varphi + 
\prt\ovl{\varphi}\pt(\mu\circ\varphi)\brp^{-1}\ .\nn
\ena
Thus, as a vector-valued $(0,1)$-form,
\be
{\mbf{\mu}}^{\varphi}\ =\ d\bz\pt\mu^{\varphi}\pt\prt\ =\ 
\blp \fbprt\varphi + \fbprt\ovl{\varphi}\pt(\mu\circ\varphi)
\brp \pt \blp \prt\varphi +
\prt\ovl{\varphi}\pt(\mu\circ\varphi)\brp^{-1}\pt\prt\ .
\lbl{mufi}
\ee

We now consider the equivalence problem between integrable complex structures.
Consider two copies of the compact complex manifold $M$ with local background
complex coordinates $\zbz$ and let $\varphi$ be any smooth diffeomorphism
homotopic to the identity map. Given a vector-valued $(0,1)$-form,
$\mbf{\mu}$, submitted to the condition $\fbprt\mbf{\mu} - \mbf{\mu}^2 = 0$, 
let $\ZBZ$ be the local complex coordinates at the point
$(w,\ovl{w})=\varphi\zbz$, by solving the system of $n$ PDE's \rf{beltra},
one can show that the 
integrability condition is preserved by
diffeomorphisms (as will be explicitly proved in \ref{invint}
and also down below for the infinitesimal version)
\be
\fbprt\mbf{\mu} - \mbf{\mu}^2\ =\ 0 \Lra
\fbprt\mbf{\mu}^{\varphi} - (\mbf{\mu}^{\varphi})^2\ =\ 0 \ ,
\ee
and thanks to the Newlander-Nirenberg theorem there are local 
complex coordinates $Z^{\varphi}$ pertaining to the integrable complex
structure parametrized by $\mbf{\mu}^{\varphi}$ ; hence one has
\be
(\fbprt -  \mbf{\mu}^{\varphi})Z^{\varphi}\ =\ 0\ .
\lbl{sourcebeltra}
\ee
The problem is to see whether the mapping
$(Z^{\varphi},\bZ^{\varphi})\lra\ZBZ$ is indeed bi-holomorphic. If it is so,
then $\mbf{\mu}$ and $\mbf{\mu}^{\varphi}$ define equivalent integrable complex
structures which correspond to the same point $t$ in the moduli space.
The situation is the same as for Riemann surfaces, that is the mapping
$(Z^{\varphi},\bZ^{\varphi})\lra\ZBZ$ is actually bi-holomorphic,
$\mbf{\mu}$ and $\mbf{\mu}^{\varphi}$ defining equivalent integrable complex
structures. The reader is refered to \ref{equivstruct} for the details. 

\indent

The infinitesimal action given by the (graded) Lie derivative,
$s \equiv L_{\fc} = i_{\fc}\, d - d\, i_{\fc}$ and, thanks to \rf{dZ}, 
writes locally in each open set $U$, 
\be
s\,Z\ =\ L_{\fc}\, Z\ =\ i_{\fc}\, d\, Z\ =\ (c + \ovl{c}\pt\mu)\pt\lam
\equiv C \pt\lam \equiv \Upsilon\ ,
\ee
where $\fc = c\pt\prt + \ovl{c}\pt\bprt = 
\Upsilon \pt \paZ + \ovl{\Upsilon}\pt \pabZ$ is
the (smooth) Faddeev-Popov ghost vector field
expressed respectively in the holomorphic
coordinates $z$ and $Z$, and $C$ reflects a change of basis 
in the Lie algebra of diffeomorphisms leading to the holomorphic factorization
property of bidimensional conformal models \cite{BB87,Bec88,KLS91a,KLS91b}.
Here, in matrix notation, we have
\be
\lp C,\bC \rp \ =\ \lp c,\bc \rp
\lp \begin{array}{c}
I \\[1mm] \hline \\[-5mm] \mu \end{array} \right|
\left. \hs{-1.2}\begin{array}{c}
\mub \\[1mm] \hline \\[-5mm] I \end{array} \rp \ ,\cm
\lp \Upsilon,\ovl{\Upsilon}\rp\ =\ \lp C,\bC \rp 
\lp \begin{array}{c}
\lam \\[1mm] \hline \\[-5mm]  0 \end{array} \right|
\left. \hs{-1.2}\begin{array}{c}
0 \\[1mm] \hline \\[-5mm] \lamb \end{array} \rp\ ,
\lbl{C}
\ee
the latter being the BRS counterpart of eq.\rf{matdZ}.

On the one hand, the infinitesimal version of \rf{finite} can be 
expressed in terms of the nilpotent BRS operation, $s^2=0$, as
\bea
s\,  \lam &=& \fc\,\lam + (\prt c + \prt\bc\pt\mu)\pt\lam\ ,\nn\\[2mm]
s\, \mu &=& \fc\, \mu + (\bprt - \mu\pt\prt)c + (\bprt - \mu\pt\prt)\bc\pt\mu
\ , \lbl{truediff}\\[2mm]
s\,\fc &=& \shalf [\fc,\fc]\ , \cm \fc\ =\ c\pt\prt + \bc\pt\bprt \nn\ ,
\ena
where the graded bracket is that of vector fields.
Note that the variation of $\mu$ is non-linear in $\mu$.
Moreover, we know that the integrability condition \rf{integ} is
$s$-invariant as shown in \ref{invint}.

On the other hand, with respect to the integrable complex structure 
defined by $\mu$, the infinitesimal action of diffeomorphisms on the
complex coordinate $Z$ writes locally in $U$, with $s^2=0$,
\bea
&&s\, Z \equiv \fc\, Z\ =\ \Upsilon,\cm
s \Upsilon\ =\ 0,\nn\\[2mm]
&& s \,\mbf{d}+\mbf{d}\,s=0 \Lra
[s,\paZ] = - \paZ\, \fc
\lbl{sZ}
\ena
Accordingly, one has the following remarkable identities, together
with their complex conjugate,\\[-6mm]
\be
[s-\fc,\paZ]\ =\ 0\ ,\cm
s\fprt+\fprt s\ =\ \fprt\ovl{\Upsilon}\pt\pabZ - \fbprt\Upsilon\pt\paZ \ .
\ee
By computing directly $s\, d\,Z=-d\, s\, Z$, one finds,
\bea
s\, \lam &=& \prt(C\pt\lam)\ ,\nn\\[2mm]
s\,\mbf{\mu} &=& - \fbprt(C\pt\prt) + [\mbf{\mu},C\pt\prt]\ ,
\cm \mbf{\mu}\ =\ d\bz\pt\mu\pt\prt\ ,\lbl{brs}\\[2mm]
s\, (C\pt\prt) &=& \shalf [C\pt\prt,C\pt\prt]\ =\ (C\pt\prt\, C)\pt\prt\ ,\nn
\ena
where the component $C$ of the vector field $C\pt\prt$, a
smooth section of the holomorphic tangent bundle $T^{1,0}$ \cite{Kod86}, 
is defined by \rf{C}. 

The reader's attention is drawn to the fact that, 
these two sytems of infinitesimal variations for $\mbf{\mu}$ and
$\lam$, \rf{truediff} and \rf{brs}, are equivalent only upon the use of both
the integrability conditions \rf{integ} and \rf{compa}. 
The former represents the true infinitesimal action of
diffeomorphisms, while the latter, which holds only in the case of an
integrable complex structure, will govern the holomorphic factorization
property. In the integrable case, the system \rf{brs} will also
represent the infinitesimal action of diffeomorphisms, and
the integrability condition \rf{integ} is seen to be stable
under diffeomorphisms by
\be
\fbprt \mbf{\mu} - \mbf{\mu}^2\ =\ 0 \Lra 
s\, (\fbprt \mbf{\mu} - \mbf{\mu}^2) \ =\ 
[C\pt\prt,\fbprt \mbf{\mu} - \mbf{\mu}^2]\ =\ 0\ .
\ee

\indent

\sect{Fields and Lagrangians}

\indent

According to what has been done in the two dimensional case
\cite{Sto88,Laz90,KLS91a,KLS91b,BaLa93}, one can
take advantage of the complex coordinate system $Z$ pertaining to an
integrable complex structure $\mu$, in order to exhibit
some simple Lagrangians built on a $n$-dimensional complex manifold.

Before going further, let us emphasize that a classical action 
$\Gamma^{\mbox{\tiny Cl}}_0$ has to be well defined on the complex
manifold $M$, \ie
\be
\Gamma^{\mbox{\tiny Cl}}_0\ =\ \int_M dZ^{A_n}\wedge
d\bZ^{\,\ovl{A}_n}\ \mbox{\large L}_{A_n\ovl{A}_n}\ZBZ\ ,
\ee
where in the integrand a multi-index notation has been used, 
see eq.\rf{Antisym} down below, with $|A_n|=|\ovl{A}_n|=n$. 
This means that the Lagrangian density is a skew-symmetric 
covariant tensorial density under 
holomorphic change of local coordinates $Z\lra Z'(Z)$,
\be
\mbox{\large L}_{A'_n\ovl{A}'_n}\ZBZp \ =\ \left|
\frac{\prt(Z'^1,\dots,Z'^n)}{\prt(Z^1,\dots,Z^n)} \right|^{-2}
\mbox{\large L}_{A_n\ovl{A}_n}\ZBZ, 
\cm |A_n|=|A'_n|=|\ovl{A}_n|=|\ovl{A}'_n|=n.
\ee

For instance, one may ask whether analogue Lagrangians 
of the $bc$--systems as well
as a generalisation of the bosonic string can be obtained over such a
higher dimensional complex manifold and what is the role of the
complex structure.

Our purpose is first to use the holomorphic coordinates $Z$ and then
to make explicit the coupling of matter fields with the integrable
complex structure $\mu$.

The simplest higher-dimensional analogue of the bidimensional $bc$-system
\cite{KLS91a,KLS91b} is constructed by considering
a holomorphic vector bundle $E$ over $M$ and the
fields $\Psi$ as a smooth section of 
${T^{k,0}}^*\tens E$, $0\leq k\leq n-1$, and
$\Psi^*$ a smooth section of ${T^{n-k,n-1}}^*\tens E^*$.
Our choice for the $bc$ fields is rather
different from the one written in \cite{LMNS97} and reduces to the
usual $bc$-system for $n=1$. One has following well-defined action
\be
S_{bc}\ =\ \int_M \Psi^*\wedge (\fbprt + \ovl{\mbf{A}})\,\Psi ,
\lbl{bc}
\ee
where the $(0,1)$-connection form $\ovl{\mbf{A}}$ parametrizes the
holomorphic structure on $E$ provided the integrability condition, 
$\fbprt \bar{\mbf{A}} + \ovl{\mbf{A}}^2 =0$, is satisfied \cite{Kos60}.

The natural generalisation of the free bosonic string action reads
\be
S(\Phi)\ =\ \int_M \fbprt\Phi\wedge \ovl{(\fbprt\Phi)}\ ,
\lbl{string}
\ee
where, in a multi-index notation, $\Phi=\Phi_I dZ^I$ is a smooth
complex-valued $(n-1,0)$-form which can be chosen to be real only in
the case $n=1$.
One can see that it is not possible to consider scalar fields in such a
complex model unless there exists a hermitian metric $g$ on $M$.

In any case, the important point is that the above Lagrangian densities
depend locally on the integrable complex structure $\mu$. One can
directly check that the two above actions are actually local in $\mu$.
Indeed, defining the following change of field variables
$(\lam^I_J)\Phi_I = \phi_J$, with 
$(\lam^I_J)=\lam^{i_1}_{j_1}\cdots\lam^{i_{n-1}}_{j_{n-1}}$, 
according to the form degree, the action
\rf{string} reads
\be
S(\phi,\mu,\mub) = \int_M \det(I-\mu\pt\mub) 
(\ovl{\mbf{\nabla}}\phi + \imath(\mbf{\tau})\phi)\wedge
\ovl{(\ovl{\mbf{\nabla}}\phi+\imath(\mbf{\tau})\phi)},
\ee
where $\mbf{\nabla}=dz\pt\nabla$ ($\nabla$ defined in \rf{nabla}) and
$\imath(\mbf{\tau})\phi=(n-1)\tau^r\wedge\phi_r$ is a $(n-1,1)$-form given
by the inner product on the $(n-1,0)$-form $\phi=\phi_I dz^I$ with respect to
the vector-valued $(1,1)$-form 
$\mbf{\tau} = d\bz\pt(I-\mu\mub)^{-1}\wedge\fprt\mu\pt\prt$. In the course
of the computation the condition \rf{compa} on the
integrating factor has been used.

Similarly, for the $bc$-system the action takes the following local
expression in $\mu$
\be
S_{bc}(\psi,\psi^*,\mu,\mub) = \int_M \det(I-\mu\pt\mub) \psi^* \wedge\lp
\ovl{\mbf{\nabla}}+ \imath(\mbf{\tau})+
d\bz\pt(I-\mu\mub)^{-1}\pt\ovl{a}\rp\psi \ ,
\ee
with $\imath(\mbf{\tau})\psi = k\,\tau^r\wedge\psi_r$ is a $(k,1)$-form
with values (as well $\psi$) in a copy of the holomorphic vector
bundle $E$ with respect to the complex analytic coordinates $z$, and
where we have set for the components of the $(0,1)$-type connection
$A_{\bZ} = \lam^{-1}\pt (I-\mu\mub)^{-1}\pt\ovl{a}$.
In this framework, (integrable) complex structures on $E$ are
parametrized by $\mu$ and $\ovl{a}$ subject to the integrability conditions
$\ovl{\mbf{\cal L}}\mbf{\mu}= 0$ and $\ovl{\mbf{\cal L}}\ovl{\mbf{a}}
+ \ovl{\mbf{a}}^2 = 0$, respectively.
However, the obvious holomorphic factorization property in $\mu$  
for $n=1$, ($\psi^*\wedge(\fbprt - \mbf{\mu} + \ovl{\mbf{a}})\psi$ 
\cite{KLS91b}) is spoilt, and the action $S_{bc}$ strongly differs 
from the one proposed in \cite{LMNS97} which is assumed to enjoy 
that property in any dimension. 

The action of diffeomorphisms gives rise, for instance, to the following
local (in $\mu$) transformation laws on the $(n-1,0)$-form $\phi$ and on
the $bc$ fields $\psi$ and $\psi^*$,
\be
s\,\phi = \fc\,\phi + \imath(\tld{\mbf{\Omega}}) \phi\ ,\cm
s\,\psi = \fc\,\psi + \imath(\tld{\mbf{\Omega}}) \psi\ ,\cm
s\,\psi^* = \fc\,\psi^* + \imath(\tld{\mbf{\Omega}}) \psi^* + 
\imath(\ovl{\tld{\mbf{\Omega}}})\psi^* \,
\ee
where we have introduced the following vector-valued $(1,0)$-form
carrying ghost grading one 
$\tld{\mbf{\Omega}}= (\fprt c + \fprt\bc\pt\mu)\pt\prt$ which
governs the parallel transport along the ghost vector field $\fc$.

\indent

\sect{Study of diffeomorphism anomalies}

\indent

From now on, since matter fields have not been specified yet, their will be
collectively denoted by $\phi$, and together with the integrable complex 
structure $\mu$, one introduces the short-hand notation 
$f_0\equiv \{\phi,\psi,\psi^*,\mu^k_\ib,\mub^\jb_l\}$, and  
for the diffeomorphism ghost fields as well
$f_1\equiv \{c^i,\bc^{\,\ib}\}$, and $f\equiv f_0\cup f_1$.
In order to suppress the index summation and the integration over the
manifold $M$ as well, we also introduce the dual contraction by the 
bracket $<,>$, defined as
\be
<\tld{f},f>\ \equiv \int_M dz^n\wedge d\bz^n <\tld{f}\zbz,f\zbz>\ .
\ee
The diffeomorphism symmetry \rf{truediff} 
can be encoded in the following nilpotent BRS functional operator
\be
\delta_0\ =\ \sum_f <s\,f,\frac{\delta}{\delta f}> ,
\cm s\, f\zbz\ =\ \fc\,f\zbz + V(f,c,\bc)\zbz\ ,
\lbl{BRSfun}
\ee
where $V$ is a local differential polynomial.
The diffeomorphism invariance of the classical theory is stated in
a Ward identity, $\delta_0\Gamma^{\mbox{\tiny Cl}}_0[f_0]=0$, together with
the integrability condition on the $\mu$'s, which may be seen as a
local constraint,
\be
\left[ \cL_\ib,\cL_\jb \right]\ =\ 
(\cL_\ib\,\mu^k_\jb - \cL_\jb\,\mu^k_\ib)\prt_k \equiv
F^k_{\ib\,\jb}(\mu)\,\prt_k,\cm \dps{\mbox{with}}\cm
F^k_{\ib\,\jb}(\mu)\ =\ 0\ ,
\lbl{const}
\ee
with of course the complex conjugate expression.
According to the diffeomorphism invariance of the
integrability condition, one has
\be
\delta_0 F^k_{\ib\,\jb}(\mu) \ =\ 0\ =\ 
\delta_0\ovl{F}^{\,\ib}_{ij}(\mub).
\ee 
Following the spirit of \cite{Bec88}, in the vacuum sector
($\phi\equiv 0$), we can rephrase the Ward identity in terms of
local Ward operators with respect to the ghost (parameter) fields,
$c$'s, as
\bea
W_k\zbz \Gamma^{\mbox{\tiny Cl}}_0[\mu,\mub] &\equiv& \lp
\prt_k \mu^r_\ib \frac{\delta}{\delta\mu^r_\ib}
- \prt_\ib\frac{\delta}{\delta\mu^k_\ib} 
+ \prt_r\blp \mu^r_\ib\frac{\delta}{\delta\mu^k_\ib}\brp\ 
+ \right.\nn\\[-2mm]
&&{}\\
&&\left.\hs{-30}
+\ \mub_k^{\ovl{s}} \blp \prt_{\ovl{s}}
		 \mub^\jb_l\frac{\delta}{\delta\mub^\jb_l}
+ \prt_\jb \blp \mub^\jb_l
\frac{\delta}{\delta\mub^{\ovl{s}}_l} \brp\brp
+\ovl{F}^{\,\jb}_{kl}(\mub)\frac{\delta}{\delta\mub^\jb_l} \rp\zbz \
\Gamma^{\mbox{\tiny Cl}}_0[\mu,\mub] \ =\ 0,\nn
\ena
together with the complex conjugate expression 
$\ovl{W}_\jb\Gamma^{\mbox{\tiny Cl}}_0[\mu,\mub] =0$.
One can directly check that the combination
\bea
\blp W_k - \mub^{\ovl{l}}_k\ovl{W}_{\ovl{l}}\brp \zbz
\Gamma^{\mbox{\tiny Cl}}_0[\mu,\mub] &=& \lp
\ovl{F}_{kl}^\jb(\mub)\frac{\delta}{\delta\mub^\jb_l}
- \mub^{\ovl{l}}_k F^r_{\ovl{l}\,\ib}(\mu)
\frac{\delta}{\delta\mu^r_\ib} + \right.\nn\\[-2mm]
&&{}\\
&&\hs{-60}\left.
+\ (I-\mub\pt\mu)^t_k \blp  \prt_r \mu^r_\ib 
\frac{\delta}{\delta\mu^t_\ib}
- \prt_\ib\frac{\delta}{\delta\mu^t_\ib} 
+ \mu^r_\ib \prt_r \frac{\delta}{\delta\mu^t_\ib}
+ \prt_t\mu^r_\ib \frac{\delta}{\delta\mu^r_\ib}\brp \rp \zbz
\Gamma^{\mbox{\tiny Cl}}_0[\mu,\mub]\ =\ 0\ ,\nn
\ena
does no longer depend on the $\mub$'s if the integrability conditions
$F(\mu\zbz)=\ovl{F}(\mub\zbz)=0$ are taken into account. With these
constraints one obtains a new local Ward operator
\bea
{\cal W}_k\zbz \Gamma^{\mbox{\tiny Cl}}_0[\mu,\mub] &\equiv&
\blp \left[(I-\mub\pt\mu)^{-1}\right]^t_k
\lp W_t - \mub^{\ovl{l}}_t \ovl{W}_{\ovl{l}}\rp\brp\zbz
\Gamma^{\mbox{\tiny Cl}}_0[\mu,\mub] \nn\\[2mm]
&=& \lp \prt_r \mu^r_\ib \frac{\delta}{\delta\mu^k_\ib}
- \prt_\ib\frac{\delta}{\delta\mu^k_\ib} 
+ \mu^r_\ib \prt_r \frac{\delta}{\delta\mu^k_\ib}
+ \prt_k\mu^r_\ib \frac{\delta}{\delta\mu^r_\ib} \rp\zbz 
\Gamma^{\mbox{\tiny Cl}}_0[\mu,\mub]\ , \lbl{fact}
\ena
which reflects the linear change of parameter $C^k=c^k+\mu^k_\ib\bc^{\,\ib}$ 
corresponding to the the variation \rf{brs} of $\mu$. Recall that the
latter is equivalent to the Lie derivative of $\mu$ only if the integrability 
condition is considered.

On the physical side, the meaning of the geometrical quantity 
$\Theta^\ib_k\zbz \equiv\dps{
\frac{\delta\Gamma^{\mbox{\tiny Cl}}_0}{\delta\mu^k_\ib\zbz}}$ 
is still obscure. It does not correspond to the energy-momentum tensor
as in the bidimensional case since the equivalence between conformal
and integrable complex structures does not hold anymore in higher dimensions.
Moreover, the $\mu$'s which serve as classical sources for the
components of $\Theta$ are not all independent due to the
integrability condition. The question whether the latter induces some
constraints on the theory is not yet under control.
Nevertheless, one may write down the classical Ward identity \rf{fact}
in the vacuum sector with respect to the ghost $C$, (the tree level)
\be
\lp \cL_\ib\zbz\, \frac{\delta}{\delta\mu^k_\ib\zbz} - 
\prt_r \mu^r_\ib\zbz \frac{\delta}{\delta\mu^k_\ib\zbz}
- \prt_k\mu^r_\ib\zbz \frac{\delta}{\delta\mu^r_\ib\zbz} \rp 
\Gamma^{\mbox{\tiny Cl}}_0[\mu,\mub]\ =\ 0.
\lbl{Wid}
\ee
Geometrically, going to the $\ZBZ$ complex coordinates, 
$\Theta$ behaves like a $(T^{1,0})^* \tens (T^{0,1})$-valued density
according to the following transformation law
\be
\TH\ZBZ \equiv \frac{1}{{\cal J}}\lam^{-1}\pt\Theta\zbz
\pt(I-\mu\pt\mub)\pt\lamb\ ,
\lbl{TH}
\ee
where ${\cal J}$ is the Jacobian defined in \rf{Jac}.
In terms of the $\ZBZ$ complex coordinates, the classical Ward identities 
\rf{Wid} translate into the vanishing of the following $n$ divergences, 
(see \ref{divfree} for some details),
\be
\prt_{Z^\rb}\, \TH^\rb_k = 0\ ,\cm k=1,\dots,n.
\lbl{ZWid}
\ee
Remark that for $n=1$, \rf{TH} reduces to the 
transformation law of a quadratic differential \cite{Laz90}.

\indent

We now introduce external fields 
$\beta\equiv\beta_{-1}\cup\beta_{-2}\equiv
\{\gamma,\eta_k^\ib,\ovl{\eta}_\jb^l\}\cup\{\zeta_i,\ovl{\zeta}_\ib\}$,
with an assigned (negative) ghost grading respectively coupled to the BRS
variation of the fields
$f\equiv \{\phi,\mu^k_\ib,\mub^\jb_l,c^i,\bc^\ib\}$,
in order to have a $\Phi$-$\Pi$ neutral action, written in a
short-hand notation as
\be
\Gamma^{\mbox{\tiny Cl}}_1[\beta,f]\ =\ \sum_{f} <\beta,s\,f> .
\ee
One has $\delta_0 \Gamma^{\mbox{\tiny Cl}}_1[\beta,f] = 0$ thanks to $s^2=0$.
Within the enlarged classical action,
$\Gamma^{\mbox{\tiny Cl}}[\beta,f]\ =\ \Gamma^{\mbox{\tiny Cl}}_0[f_0]+
\Gamma^{\mbox{\tiny Cl}}_1[\beta,f]$, the BRS functional operator writes
\be
\delta_0\ =\ \sum_{f}
<\frac{\delta\Gamma^{\mbox{\tiny Cl}}}{\delta\beta},\frac{\delta}{\delta f}> 
\ee
and the diffeomorphism invariance of $\Gamma^{\mbox{\tiny Cl}}$
simplifies into the so-called Slavnov identity \cite{PR81}
\be
\delta_0\Gamma^{\mbox{\tiny Cl}}[\beta,f]\ =\ 0.
\ee
In order to preserve the classical diffeomorphism invariance on the functional
$\Gamma^{\mbox{\tiny Cl}}$, the BRS operation can be extended to the 
classical sources $\beta$, either trivially
by saying that these external fields are $s$-invariant, or by defining
\bea
s\beta_{-1}\zbz &=& 
\frac{\delta\Gamma^{\mbox{\tiny Cl}}[\beta,f]}{\delta f_0\zbz}\ =\ 
\frac{\delta\Gamma^{\mbox{\tiny Cl}}_0[f_0]}{\delta f_0\zbz}
+ \fc\, \beta_{-1}\zbz + \tld{V}(\beta_{-1},f_1)\zbz, \nn\\[-3mm]
&{}&\lbl{ssource}\\
s\beta_{-2}\zbz &=& 
\frac{\delta\Gamma^{\mbox{\tiny Cl}}[\beta,f]}{\delta f_1\zbz}\ =\ 
\fc\, \beta_{-2}\zbz + \tld{V}(\beta_{-2},f_1)\zbz + 
X(f_0,\beta_{-1})\zbz\ ,\nn
\ena
where $\tld{V}$ is the formal dual of the differential operator $V$
given in \rf{BRSfun}, and $X$ is the local differential polynomial
\be
X_i(f_0,\beta_{-1})\zbz\ =\
\frac{\delta}{\delta c^i\zbz} \sum_{f_0} <\beta_{-1},sf_0>\ ,
\ee
together with the complex conjugate expression. With these variations we
can define the following nilpotent functional differential operator
\be
\delta\ =\ \delta_0 + 
\sum_{\beta} <s\beta,\frac{\delta}{\delta\beta}>\ ,
\ee
which corresponds to the linearized Slavnov operator acting on the
space of local functionals in both the fields $f$ and the sources $\beta$.
Finding a quantum extension $\Gamma[\beta,f]$ of the classical functional
$\Gamma^{\mbox{\tiny Cl}}$ such that $\delta \Gamma = 0$ amounts to
studing the $\delta_0$-cohomology in the space of local functionals
with respect to the ghost grading (the $\Phi\Pi$ charge). In the
lagrangian field theory language, a quantum extension will
exist if the cohomology is trivial at the ghost grading one, \ie there
is no anomaly.
It is now custumary \cite{Dix76,Ba88} to translate this cohomological 
problem into the space of local polynomials in the fields and sources and their
derivatives spanned by the infinite set of local independent variables
of the type
\bea
&& \prt_I\prt_{\ovl{J}}\, \chi\zbz, \cm 
\chi \equiv \{\phi,\mu,\mub,c,\bc,\gamma,\eta,\ovl{\eta},\zeta,\ovl{\zeta}\},
\cm |I|\geq 0,\ |\ovl{J}|\geq 0,\nn\\[3mm]
&& \prt_I\equiv (\prt_1)^{i_1}\dots(\prt_n)^{i_n},\cm I\ =\ (i_1,\dots,i_n),
\cm |I|\ =\ i_1 + \cdots + i_n, \nn\\
&& \prt_{\ovl{J}}\equiv (\bprt_1)^{j_1} \dots (\bprt_n)^{j_n}, \cm
\ovl{J}\ =\ (j_1,\dots,j_n),\cm |\ovl{J}|\ =\ j_1+\cdots+j_n\ ,\nn\\
&& \delta_I^{M+N}\frac{I!}{M!N!}\equiv \prod_{k=1}^{n} 
\delta^{m_k+n_k}_{i_k}\frac{i_k!}{m_k!n_k!}
\lbl{multiI}\\
&& I+1_l\ =\ (i_1,\dots,i_l+1,\dots,i_n),\cm
\ovl{J}+1_{\ovl{l}}\ =\ (j_1,\dots,j_l+1,\dots,j_n),\nn
\ena
where we have conveniently used a multi-index notation and $|I|$
denotes the length of the multi-index $I$. The space of local
polynomials in the fields and their derivatives can be endowed with the
structure of a Fock space.
The interchange between the two systems of local complex coordinates
$\zbz$ and $\ZBZ$, can be taken into account. According to a previous
work \cite{BaLa93}, by considering the coordinates $Z$ and $\bZ$ as 
independant variables but non local in $\mu$, we enlarge the set 
$\chi\equiv f\cup\beta$ of fields. The local cohomology will be that of the
following differential operator
\bea
\delta &=& <sZ\zbz, \DER{Z}> + <s\bZ\zbz, \DER{\bZ}> \nn\\
&&+\ \sum_{|I|,|\ovl{J}|\geq 0} \lp \sum_{\chi}
<\prt_I\prt_{\ovl{J}}\, s \chi\zbz, \DER{\prt_I\prt_{\ovl{J}}\, \chi}>
\right. \lbl{brslocal} \\
&&\left. +\ 
<\prt_I\prt_{\ovl{J}}\, s\lam\zbz, \DER{\prt_I\prt_{\ovl{J}}\,\lam}>
+ <\prt_I\prt_{\ovl{J}}\, s\lamb\zbz, \DER{\prt_I\prt_{\ovl{J}}\,\lamb}>\rp\ ,
\nn
\ena
which turns out to be nilpotent $\delta^2=0$ upon using both the integrability
conditions \rf{compa} and \rf{integ}. Note that the non local part in
the $\mu$'s has been isolated. 

For a given $\Phi$-$\Pi$ charge $p$, solutions will be obtained modulo
total derivatives, and we get the following descent equations
\be
\left\{ \begin{array}{l}
\delta \Delta_{2n}^{p}\zbz + d \Delta_{2n-1}^{p+1} \zbz =0 \\[2mm]
\delta \Delta_{2n-1}^{p+1}\zbz + d \Delta_{2n-2}^{p+2} \zbz =0 \\[2mm]
\hs{20}\vdots \\
\delta \Delta_{2n-r}^{p+r}\zbz + d \Delta_{2n-r-1}^{p+r+1} \zbz =0 \\
\hs{20}\vdots \\
\delta \Delta_{1}^{p+2n-1}\zbz + d \Delta_{0}^{p+2n} \zbz =0 \\[2mm]
\delta \Delta_{0}^{p+2n}\zbz =0 
\end{array} \right.                                 
\lbl{descenteqs}   
\ee                                        
where the lower index will label the Poincar\'e form degree 
and the upper index denotes the ghost grading.
The last of these equation has general solution 
\be
\Delta_{0}^{p+2n}\zbz \ = \ \Delta_{0}^{p+2n, \natural}\zbz +
\delta\widehat{\Delta}_{0}^{p+2n-1}\zbz
\lbl{soldelta}
\ee
where $\Delta_{0}^{p+2n, \natural}\zbz$ is an element of the 
$\delta$-cohomology space in the space of local functions.
It is readily seen \cite{Ba88,BaLa93} that
\be
\mbf{d}\ =\ dz^k \left\{ \delta,\DER{c^k} \right\} + 
d\bz^k \left\{ \delta,\DER{\bc^k} \right\}\ ,
\ee
which allows one to define the derivative operator on the local
polynomials as a formal derivative
\bea
\leqa{\prt_i\ =\ \left\{ \delta,\DER{c^i} \right\}\ =\
<\lam_i\zbz, \DER{Z}> + <(\mub\pt\lamb)_i\zbz,\DER{\bZ}> }\nn\\
&&+ \sum_{|I|,|\ovl{J}|\geq 0} \lp 
<\prt_{I+1_i}\prt_{\ovl{J}}\, \lam\zbz, \DER{\prt_I\prt_{\ovl{J}}\,\lam}>
+ <\prt_{I+1_i}\prt_{\ovl{J}}\, \lamb\zbz, \DER{\prt_I\prt_{\ovl{J}}\,\lamb}>
\right. \nn\\
&&\left. +\ \sum_{\chi}
<\prt_{I+1_i}\prt_{\ovl{J}}\, \chi\zbz, \DER{\prt_I\prt_{\ovl{J}}\, \chi}> \rp\ ,
\lbl{derivative} \ena
and one gets $\bprt_i$ by complex conjugation.

Going up through the system of descent equations \rf{descenteqs} 
we subtitute Eq\rf{soldelta} and obtain
\be
\delta\left[ \Delta_{1}^{p+2n-1}\zbz 
+\DERR{\Delta_{0}^{p+2n,\natural}}{c^i} dz^i 
+\DERR{\Delta_{0}^{p+2n,\natural}}{\bc^\ib} d\bz^\ib  
+ \mbf{d} \widehat{\Delta}_{0}^{p+2n-1}\zbz \right] = 0 \ .
\ee
This implies that
\bea
\Delta_{1}^{p+2n-1}\zbz + 
\DERR{\Delta_{0}^{p+2n,\natural}}{c^i} dz^i 
+\DERR{\Delta_{0}^{p+2n,\natural}}{\bc^\ib} d\bz^\ib + 
\mbf{d} \widehat{\Delta}_{0}^{p+2n-1}\zbz  
&=&\nn\\[2mm]
&&\hs{-80}=\ \Delta_1^{p+2n-1,\natural} 
\zbz + \delta\widehat{\Delta}_{1}^{p+2n-2}\zbz 
\ena
and proceeding step by step, we reach the top equation of \rf{descenteqs}
which gives the general solution of the $\delta$-cohomology in the
sector of ghost grading $p$ \cite{Ba88,BaLa93}
\bea
\Delta_{2n}^{p}\zbz &=& \sum_{r,s=1}^n
\frac{\prt^r}{\prt c^{A_r}\zbz} 
\frac{\prt^s}{\prt \bc^{\ovl{A}_s}\zbz}\Delta_{2n-r-s}^{p+r+s,\natural} \zbz 
\, dz^{A_r}\wedge d\bz^{\ovl{A}_s} \nn\\[2mm]
&&+\ \Delta_{2n}^{p,\natural}\zbz + \delta \widehat{\Delta}_{2n}^{p-1} 
\zbz + \mbf{d} \widehat{\Delta}_{2n-1}^{p}\zbz \ ,
\lbl{solutionnatural}
\ena
where for the sake of compactness an increasing ordered multi-index
notation has been used,
\be
A_r= (a_1,\dots,a_r)\ ,\ a_1<\cdots<a_r\ ,\
dz^{A_r}=dz^{a_1}\wedge\cdots\wedge dz^{a_r}\ , \
\frac{\prt^r}{\prt c^{A_r}}= 
\frac{\prt^r}{\prt c^{a_1}\cdots \prt c^{a_r}}\ ,
\lbl{Antisym}
\ee
with of course the complex conjugate expressions.
Equation \rf{solutionnatural} links the elements of the diff-mod $d$ 
cohomology to those of the local functional cohomology, once the 
$\Delta_{2n-r-s}^{p+r+s,\natural}$, $r,s=0,\dots,n$ are known as
$\delta$-cocycles. 

We are interested in finding possible anomalies in the quantum field
theoretic sense, hence with $p=1$ in eq.\rf{solutionnatural}.
According to our previous analysis in \cite{BaLa93}, it turns out that the only
relevant tensorial index $2n-r-s$ is of scalar type, namely for
$r=s=n$. The problem reduces into solving the cocycle equation
\be
\delta\Delta_0^{2n+1,\natural} \zbz  \ =\ 0\ .
\lbl{cocycle}
\ee
Following the line of Refs \cite{Ba88,BaLa93,BaLa95}
we decompose the general $\Delta_0^{p,\natural}$, for
$p\geq 2n$, with respect to its underived ghosts content, namely 
\be
\Delta_0^{p,\natural} \zbz\ =\ \sum_{r,s=0}^n 
c^{A_r}\zbz\,\bc^{\,\ovl{A}_s}\zbz\, {\cal D}_{A_r\ovl{A}_s}^{p-r-s}\zbz\ ,
\lbl{decomposition1}
\ee
where according to the summation on all possible 
ordered multi-indices \rf{Antisym}, 
the ${\cal D}$'s are independent skew-symmetric tensors containing only 
derivatives of the ghost fields. This canonical decomposition
identifies independent sectors according to the ghost grading.
Inserting the decomposition \rf{decomposition1} into \rf{cocycle}, 
and introducing the nilpotent operator
\be
\hdelta =\ \delta - \lp c^i\zbz \prt_i +c^\ib\zbz \prt_\ib \rp \ , \cm
\hdelta^2\ =\ 0\ ,
\lbl{hdel}
\ee
the cocycle condition can be canonically decomposed with respect to
the underived ghost fields onto each ghost sector, and find the 
following chain of $2^{2n}$ equations
\be
\left\{ \begin{array}{l}
\hdelta {\cal D}^p \zbz\ =\ 0 \\[2mm]
-\hdelta{\cal D}^{p-1}_\ib\zbz + \prt_\ib\bc^\jb
{\cal D}^{p-1}_\jb\zbz + \prt_\ib c^j
{\cal D}^{p-1}_j\zbz +\prt_\ib{\cal D}^p \zbz\ =\ 0 \\[2mm]
-\hdelta{{\cal D}}^{p-1}_i\zbz + \prt_ic^j
{\cal D}^{p-1}_j\zbz + \prt_i \bc^\jb
{\cal D}^{p-1}_\jb \zbz + \prt_i{\cal D}^{p}\zbz\ =\ 0 \\
\hs{20}\vdots \\
(-)^{k+l}\,\hdelta{\cal D}^{p-k-l}_{A_k\ovl{A}_l}\zbz 
+ \dps{\epsilon^{A_ri}_{A_k} \left( 
\sum_{t=1}^k (-)^{t-1+l}\, \prt_i c^{a_t} 
{\cal D}^{p-k-l}_{A_{r+1_t}\ovl{A}_l}\zbz \right.} \\[2mm]
+ \dps{\left. \sum_{t=1}^{l+1} (-)^{k+l+t}\prt_i\bc^{\,\ovl{a}_t}
\,\delta^{\ovl{A}_s}_{\ovl{A}_l}\, {\cal D}^{p-k-l}_{A_r\ovl{A}_{s+1_t}}\zbz 
+ (-)^{k-1} \prt_i {\cal D}^{p-k-l-1}_{A_r\ovl{A}_s}\zbz \right)}\\[2mm]
+\ \dps{\epsilon^{\ovl{A}_s\ib}_{\ovl{A}_l} \left(
\sum_{t=1}^{k+1} (-)^{t-1}\, \prt_{\,\ib} c^{a_t}\, \delta^{A_r}_{A_k}\,
{\cal D}^{p-k-l}_{A_{r+1_t}\ovl{A}_s}\zbz + \sum_{t=1}^l (-)^{k+t-1}\,
\prt_{\,\ib} \bc^{\,\ovl{a}_t} {\cal D}^{p-k-l}_{A_k\ovl{A}_{s+1_t}}\zbz 
\right.}\\[5mm]
+ \left. (-)^{k+l-1}\, \prt_{\,\ib} {\cal D}^{p-k-l-1}_{A_k\ovl{A}_s}\zbz
\right)\ =\ 0\\
\hs{20}\vdots \\
\lp \hdelta - \prt_i c^i\zbz -\prt_\ib \bc^\ib \zbz \rp
{\cal D}^{p-2n}_{A_n\ovl{A}_n}\zbz + 
(-1)^{n-1}\epsilon^{A_{n-1}r}_{A_n} \prt_r 
{\cal D}^{p-2n+1}_{A_{n-1}\ovl{A}_n}\zbz \\[3mm]
\hs{60}-\ \epsilon^{\ovl{A}_{n-1}\jb}_{\ovl{A}_n} \prt_\jb
{\cal D}^{p-2n+1}_{A_n\ovl{A}_{n-1}}\zbz \ =\ 0\ ,
\end{array} \right.                                    
\lbl{consistency1}
\ee                                        
where $\epsilon^{A_ri}_{A_k}$ is the generalized Kronecker
skew-symmetric tensor with respect to the multi-indices, and $A_{r+1_t}$ means
$a_1<\cdots<a_t<\cdots<a_r$ for a given $A_r$. In the above formulaes
summation on repeated multi-indices must be performed and 
$\delta^{\ovl{A}_s}_{\ovl{A}_l}$ equals 1 if the ordered multi-indices
are identical and 0 otherwise.

Since the $\delta$-cohomology is translated in terms of the
$\hdelta$-cohomology, the first step is to solve the top equation of 
the system \rf{consistency1}. 
Also, note that the $\hdelta$ operator will act on the space of local
polynomials in the following extended set of fields
$\chi\equiv f\cup\beta\cup\{\lam,\ovl{\lam}\}$.
Solving the $\hdelta$-cohomology is the
technical part of the paper and in \ref{coho} the following theorem
will be established.

\vskip .5cm\noindent
{\bf Theorem.} {\em Define in matrix notation the $n^2$
local expressions of ghost grading one,
\be
\tld{\Omega}\zbz = (\prt c + \prt \bc\pt\mu) \zbz =
(\hdelta\lam\pt\lam^{-1})\zbz ,
\lbl{omegatld}
\ee
and note the $\hdelta$-coboundary}
\be
\tr\,\tld{\Omega}\zbz\ =\ \hdelta\ln\det\lam\zbz .
\ee
{\em Then, in the scalar sector of the 
space of analytic (local) functions in the fields and their derivatives, 
the ghost sectors with grading $p$
of the ${\hdelta}$-cohomology are generated by the following 
non-trivial cocycles}
\be
\mbox{\nms for even}\ p\ :\
\tr\,\tld{\Omega}^{2r+1}\zbz\,\tr\,\ovl{\tld{\Omega}}{}^{2s+1}\zbz,\cm
\mbox{\nms for odd}\ p,\ :\ 
\tr\,\tld{\Omega}^{2k+1}\zbz,
\ee
{\em while the cocycle}
$\tr\tld{\Omega}$ {\em can be reabsorbed by completing the set of
generators with $\ln\det\lam$ seen as an independent 
variable. The same statements hold true for the complex conjugate expressions.}

\indent

A non trivial {\em local} anomaly modulo $\mbf{d}$ will then be given by 
(cf. eq.\rf{solutionnatural})
\be
\Delta_{2n}^1\zbz\ =\ \frac{\prt^n}{\prt c^{A_n}\zbz} 
\frac{\prt^n}{\prt \bc^{\ovl{A}_n}\zbz}\Delta_0^{2n+1,\natural} \zbz 
\, dz^{A_n}\wedge d\bz^{\ovl{A}_n}\ ,
\lbl{nat}
\ena
with the {\em local} $\delta$-cocycle $\Delta_0^{2n+1,\natural}$ given by the
decomposition \rf{decomposition1} and also subject to the reality
condition $\ovl{\Delta}_{2n}^1\zbz = \Delta_{2n}^1\zbz$.

According to the definition of both $\hdelta$ and \rf{omegatld}, one
considers in matrix notation the $n^2$ variables of ghost grading one,
\bea
\Omega\zbz &=& (\delta\lam\pt\lam^{-1})\zbz\ =\ \tld{\Omega}\zbz +
(\fc\,\lam \pt\lam^{-1})\zbz,\nn\\
\tr\,\Omega\zbz &=& \delta\ln\det\lam\zbz.\nn
\ee
Unfortunately, $\Omega$ is not local in the $\mu$ fields due to the
presence of the $\lam$ terms.
One has $\delta\Omega\zbz = \Omega^2\zbz$. Moreover, setting $\Lam_i
\equiv \prt_i\lam\pt\lam^{-1}$, it is easily shown
that
\bea
\Omega\zbz &=& (C^r\Lam_r)\zbz + \prt C\zbz,\nn\\[-3mm]
&{}&\lbl{1/2loc}\\[-3mm]
\delta\ln\det\lam\zbz &=& (C^r\prt_r \ln\det\lam)\zbz + \tr(\prt C)\zbz,\nn
\ena
after the use of eq.\rf{compa} and the definition \rf{C} of the
factorized ghosts $C$. Note that the above variations contain a local
part in the $\mu$ fields.

In turns out, after a rather lengthy 
analysis based on the above theorem, that the possible
$\delta$-cocycles $\Delta_0^{2n+1,\natural}$ are not local in the $\mu$
external fields due to the presence of the non-local $\lam$ terms. The
former are listed hereafter
\be\begin{array}{ll}
\tr\,\Omega^n\tr\,\ovl{\Omega}^n\tr\Omega =
\delta(\tr\,\Omega^n\tr\,\ovl{\Omega}^n\ln\det\lam),&\cm \mbox{\nms for
odd}\ n ,\\[2mm] 
\tr\,\Omega^{2r+1}\tr\,\ovl{\Omega}^{2s+1}\tr\,\Omega =
\delta(\tr\,\Omega^{2r+1}\tr\,\ovl{\Omega}^{2s+1}\ln\det\lam),&\cm 
\mbox{\nms for even}\ n ,\ n=r+s+1,\\[2mm]
\tr\,\Omega^{2n+1} + \tr\,\ovl{\Omega}^{2n+1} \neq \delta(\cdots),&\cm
n\geq 3. \end{array}
\ee
The first two cocycles are trivial while the last non-trivial 
$\delta$-cocycle can only produce a local
cocycle of ghost grading $2n$ and not $2n+1$ as required. This shows
that the local $\delta$-cohomology modulo $\mbf{d}$
is not trivial in that sector.

Therefore, if there were an anomaly, 
the latter is a non-trivial cocycle of the $\delta$-cohomology 
modulo $\mbf{d}$. So it is not surprising that the last 
possibility to be considered is 
that the {\em local} $\delta$-cocycle $\Delta_0^{2n+1,\natural}$ 
matching with the principle of locality will appear as the variation of
a non-local expression in the $\mu$ fields according to,
\be
\Delta_0^{2n+1,\natural}\ =\ \delta\, \widehat{\Delta}^{2n}_0,
\lbl{nonloc}
\ee
where $\widehat{\Delta}^{2n}_0$ does not belong to the non-local 
$\delta$-cohomology modulo $\mbf{d}$ and will provide non-local
counterterms in the $\mu$ fields which are defined up to a total
derivative. Since the non-local generators are the $\lam$ terms and
the variations \rf{1/2loc} contain a local part, 
$\widehat{\Delta}^{2n}_0$ can be formally expanded in terms of 
$\lam$ and its (independent) derivatives. By using once more eq.\rf{compa},
it assumes in the scalar sector the following finite expansion
\be
\widehat{\Delta}^{2n}_0 = \sum_{|I|=0}^{2n} \left( 
T^I\prt_I\ln\det\lam + 
\tr \blp {\cal T}^{,I+1_i}\pt\prt_I\Lam_i\brp + \mbox{c.c.}\right) ,
\ee
where in multi-index notation \rf{multiI}
both the scalar valued, $T^I=
T^{\overbrace{\sst 1\dots 1}^{i_1}\dots\overbrace{\sst n\dots n}^{i_n}}$ 
and matrix valued, ${\cal T}^{,I+1_i}$, symmetric tensors 
are local expressions in the $\mu$
external fields with $\dim T^I=\dim{\cal T}^{,I+1_i}+1=2n-|I|$, and
$Q_{\Phi\Pi}(T^I)=Q_{\Phi\Pi}({\cal T}^{,I+1_i})=2n$.
Recalling that in a Lagrangian formulation counterterms are defined up
to a total derivative, so it is for those generated by 
$\widehat{\Delta}^{2n}_0$. Thus, discarding the explicit form of
the divergences one can write  
\bea
\widehat{\Delta}^{2n}_0 &=&\!\! \lp \sum_{|I|=0}^{2n}(-1)^{|I|}\prt_I
T^I\rp\ln\det\lam + \tr\lp\blp \sum_{|I|=0}^{2n}(-1)^{|I|}\prt_I{\cal
T}^{,I+1_i}\brp \pt \Lam_i\rp + \prt_i \Gamma^i + \mbox{c.c.}\nn\\[2mm]
&\equiv& R\, \ln\det\lam + \tr \lp {\cal R}^{,i}\pt\Lam_i\rp+ \prt_i
\Gamma^i + \mbox{c.c.}
\ena
with $\dim R = \dim {\cal R}^{,i} + 1= 2n = Q_{\Phi\Pi}(R) =
Q_{\Phi\Pi}({\cal R}^{,i})$. Substitution of
$\widehat{\Delta}^{2n}_0$ in eq.\rf{nat}, together
with $[\DER{c^\al},\prt_{\beta}]=0$, provides the anomaly up to a
total derivative, one can thus discard the divergences.
The $\delta$-variation of $\widehat{\Delta}^{2n}_0$ writes
\bea
\delta\widehat{\Delta}^{2n}_0\! &=&\!  \underbrace{R\,\tr\prt C + 
\tr\lp{\cal R}^{,i}\prt_i\pt\prt C\rp}_{local\ part}  
+\, \delta R\,\ln\det\lam + R\,C^r\prt_r\ln\det\lam \nn\\
&&+\ \tr\lp \delta{\cal R}^{,i}\pt\Lam_i
+{\cal R}^{,i}\pt (C^r\prt_r\Lam_i + \prt_i C^r\Lam_r + 
[\prt C,\Lam_i])\rp + \mbox{c.c.},
\ena
and by requiring the vanishing of the non-local part one gets 
\be (a)\left\{\! \begin{array}{l} 
R\, C^r = 0,\\
{\cal R}^{,i}C^r = 0, \end{array}\ \forall r=1,\dots,n,\cm
(b)\right\{\! \begin{array}{l}\delta R = 0, \\
\delta{\cal R}^{k,i}_l = {\cal R}^{k,i}_s \prt_l C^s - {\cal
R}^{k,s}_l\prt_sC^i - {\cal R}^{s,i}_l\prt_s C^k\ . \end{array}
\lbl{nonloc=0}
\ee
Remark that ${\cal R}^{k,i}_l$ tranforms in a rigid manner.
Eqs.\rf{nonloc=0}(a) are solved by
\be (a)\left\{\!\begin{array}{l}
R\ =\ \Xi_{A_n} C^{A_n},\\
{\cal R}^{,i}\ =\ \Sigma^{,i}_{A_n}C^{A_n},\end{array}\right.\cm\cm
(b)\ \frac{\prt \Xi_{A_n}}{\prt C^r} = 
 \frac{\prt \Sigma^{,i}_{A_n}}{\prt C^r} = 0,\ 
\forall r=1,\dots,n,
\lbl{locsol}
\ee
with $C^{A_n}\equiv C^1C^2\dots C^n$, $\delta C^{A_n}= - \tr(\prt
C)\,C^{A_n}$, and 
$Q_{\Phi\Pi}(\Xi_{A_n}) = Q_{\Phi\Pi}(\Sigma^{,i}_{A_n}) = n$.
The local cocycle \rf{nonloc} then reads
\be
\Delta_0^{2n+1,\natural}\ =\ C^{A_n} \lp \Xi_{A_n} \prt_r C^r + 
\Sigma^{k,i}_{l,A_n} \prt_i \prt_k C^l \rp + \prt_r\Gamma^r_{loc} +
\mbox{c.c.}
\lbl{loco}
\ee
A direct substitution of \rf{locsol}(a) into Eqs.\rf{nonloc=0}(b)
gives,
\be
\lp \delta\Xi_{A_n} - \prt_r C^r \Xi_{A_n} \rp C^{A_n} = 0 \Lra 
\delta \Xi_{A_n} =  \tr(\prt C)\,\Xi_{A_n} + C^r \xi_{r,A_n} , 
\lbl{consistC0}
\ee
with $Q_{\Phi\Pi}(\xi_{r,A_n}) = n$, and futhermore,
\be
\delta \Sigma^{k,i}_{l,A_n}\ =\ \tr(\prt C)\, \Sigma^{k,i}_{l,A_n} +
\prt_l C^s \Sigma^{k,i}_{s,A_n} - \prt_s C^i \Sigma^{k,s}_{l,A_n}
- \prt_s C^k \Sigma^{s,i}_{l,A_n} + C^r \sigma^{k,i}_{r,l,A_n}\ ,
\lbl{consistC1}
\ee
with $Q_{\Phi\Pi}(\sigma^{k,i}_{r,l,A_n}) = n$.
Obviously, due to \rf{locsol}(b), the $C$
dependence in both eqs.\rf{consistC0} and \rf{consistC1} comes
from the underived ghost content of the BRS operator and thus 
\be
\frac{\prt \xi_{A_n}}{\prt C^r} = 
\frac{\prt \sigma^{k,i}_{l,A_n}}{\prt C^r} = 0,
\cm \forall r=1,\dots,n.
\lbl{C-locsol}
\ee
The terms $\Xi_{A_n}$, $\xi_{r,A_n}$, $\Sigma^{k,i}_{l,A_n}$ and
$\sigma^{k,i}_{r,l,A_n}$ are local in the fields
$\{\phi,\mu,\mub,c,\bc\}$ and their explicit form has to be separately
discussed according to the presence of the $\phi$ matter fields. 
The same considerations hold for the complex conjugate expressions. 

\subsect{Vacuum Anomalies}

\indent

In the vacuum sector, it is readily seen that ${\dps \prt_r =
\biggl\{\frac{\prt}{\prt C^r},\delta\biggr\}}$. This easily implies 
together with eqs.\rf{locsol}(b), and \rf{C-locsol} that
$\xi_{r,A_n} = \prt_r \Xi_{A_n}$, and 
$\sigma^{k,i}_{r,l,A_n}= \prt_r
\Sigma^{k,i}_{l,A_n}$. Eqs.\rf{consistC0} and \rf{consistC1} rewrite
respectively as 
\be
\delta \Xi_{A_n} = \prt_r(C^r\Xi_{A_n}),\cm
\delta \Sigma^{k,i}_{l,A_n} = \prt_r(C^r\Sigma^{k,i}_{l,A_n}) +
\prt_l C^s \Sigma^{k,i}_{s,A_n} - \prt_s C^i \Sigma^{k,s}_{l,A_n}
- \prt_s C^k \Sigma^{s,i}_{l,A_n},
\ee
showing that $\Xi_{A_n}$ and $\Sigma^{k,i}_{l,A_n}$ are respectively 
a scalar and a tensorial densities. The inhomogeneous term
in the $\delta$-variation of $\mu$ eliminates the possible $\mu$ dependence of
both $\Xi_{A_n}$ and $\Sigma^{k,i}_{s,A_n}$; one has to figure out 
the latter in terms of local expressions in the derivated ghosts $C$'s only.
According to both the power counting index $2n$ and the ghost grading $n$, 
it turns out that the most general expressions are
labelled by the set of permutations $\pi$ of the integers
$\{1,\dots,n\}$. Thus there are $n!$ scalar densities of the following type 
\be
\Xi_{{A_n},(\pi)} = \epsilon^{a_1\cdots a_n}\prt_{a_1}\prt_{i_1}C^{l_1}
\cdots\prt_{a_n}\prt_{i_n}C^{l_n} 
\delta^{i_1}_{l_{\pi(1)}}\cdots \delta^{i_n}_{l_{\pi(n)}}\ ,
\ee
the $\pi$'s cover all the possible contractions. 
Saturation with the product $C^{A_n}$ of the $C$'s provides the solutions
\rf{locsol}. The latter are numerized according to the decompositions 
of the integer $n$ into sums of intergers independently to the order.
Let $p(n)$ denote such a decomposition. The independent solutions of 
\rf{locsol} can be seen as skewsymmetrizations of products of already 
skewsymmetric tensors of maximal rank,
\be
R_{(p(n))} = \frac{(-1)^{\frac{n(n-1)}{2}}}{k_1!\cdots k_n!}\,
\tr\lp V^{k_1}\rp\cdots \tr\lp V^{k_n}\rp \ ,
\ee
for a given partition $p(n)$ of the integer $n$ with 
$k_1 + \cdots + k_n = n$, $k_1\leq \cdots \leq k_n$ and
where in matrix notation the $n^2$ ghost graded 2 quantities 
$V_i^l\equiv C^r\prt_r\prt_i C^l$ rigidly transforms as
\be
\delta V = \left[\prt C, V \right]\ ,\cm V\equiv C\pt\prt \lp \prt C \rp ,
\ee
a transformation law very similar to that of the adjoint of
$gl(n,\mbf{C})$ with matrix parameter $\prt C$.
The first contributions $R_{(p(n))}\,\prt_rC^r$ to the cocycle \rf{loco} are
developed with respect to the monomial $\prt_rC^r$ while the second
are developed with respect to the monomials $\prt_i \prt_k C^l$. It is
easily seen that $\Sigma^{k,i}_{s,A_n} \prt_k\prt_i C^s$ is a scalar
density of power counting index $2n+1$ and ghost grading $n+1$, so that
$\Sigma^{k,i}_{s,A_n} \prt_k\prt_i C^s = \Xi_{A_n}\tr(\prt C)$.
One has, as diffeomorphism cocycles labelled by all the partitions of $n$,
\be
R_{p(n)} \tr\lp\prt C\rp = {\cal R}^{k,i}_{l,p(n)} \prt_k\prt_i C^l\ .
\lbl{GF}
\ee
Thus, in the vacuum sector, the $\delta$-cocycle modulo $\mbf{d}$ 
\rf{loco} reduces to the following linear combination
\be
\Delta_0^{2n+1,\natural}\ =\ \sum_{p(n)} A_{p(n)} R_{(p(n))}\,\tr\prt C
\lbl{anomcoc}
\ee
where the $A_{p(n)}$'s are complex numbers depending on the field
content of the model. The cocycles \rf{GF} are higher 
dimensional analogues of the well-known Gel'fand-Fuchs cocycle 
\cite{Fucks86}. For complex dimension $n=1$, one recovers the usual
Gel'fand-Fuchss cocycle $C\prt C \prt^2C$ \cite{BaLa93}.

Inserting the cocycle \rf{anomcoc} into the computational formulae
\rf{nat}, one finds the local expression of (U-V part)
the vacuum anomaly modulo $\mbf{d}$
\bea
\Delta_{2n}^1\!&=&\! (-1)^{\frac{n(n-1)}{2}} 
\sum_{p(n)} A_{p(n)}\,\frac{1}{k_1!\cdots k_n!}\, \grt{[}
\tr\lp \prt C \rp \tr\lp (\prt{\cal V})^{k_1} \rp \cdots
\tr\lp (\prt{\cal V})^{k_n}\rp  \nn\\
&& +\ \tr\lp \prt{\cal V}\rp \sum_{l=1}^n k_l\  
\tr\lp (\prt{\cal V})^{k_1} \rp \cdots
\tr\lp (\prt{\cal V})^{k_l - 1} \prt C \rp \cdots
\tr\lp (\prt{\cal V})^{k_n}\rp \grt{]}\ ,
\lbl{finam}
\ena
where the sum is performed over all the partitions $p(n)$ 
of the integer $n$, $k_1 + \cdots + k_n = n$, with 
$k_1\leq \cdots \leq k_n$, and
where we have introduced the $n^2$ vector-valued $(1,1)$-forms 
\be
{\cal V}\equiv\fprt(d\bz\pt\mu) = \lp dz\pt\frac{\prt}{\prt c} \rp 
\lp d\bz\pt\frac{\prt}{\prt\bc}\rp V\ ,
\ee
fulfilling the following Bianchi-like identities
\be
\fprt {\cal V} = 0,\cm \fbprt{\cal V} + [{\cal V},\mbf{\mu}] = 0,
\ee
thanks to the integrability condition \rf{integ}. In the $n=1$ case,
the well-known holomorphically split anomaly $\prt C\prt^2\mu$
\cite{BB87,Bec88} is recovered.

Note that there is a very particular linear combination of the
anomaly \rf{finam} given by the component of degree $2n+2$ of the Todd class 
according to formula (5.19) of \cite{LMNS97}. However, formula (1.11)
or (8.5) of the quoted reference does not match as a diffeomorphism
anomaly in the spirit of our construction in the sense that it 
is not a diffeomorphism $\delta$-cocycle whereas each term in the 
summand \rf{finam} is. 

In concluding, since the same argument holds for the complex
conjugate part, the anomalous Ward id's \rf{Wid} are holomorphically split 
at the quantum level, similarly to the bidimensional case, it is a
strong indication that the vacuum functional turns out to be as well 
holomorphically factorized in the Beltrami differential
parametrizing an integrable complex structure. The anomalous Ward
id's correspond to the non-vanishing of the divergence
\rf{ZWid} at the quantum level, a similar phenomenon to the
bidimensional case, \cite{Laz90}.

A non-local functional $\Gamma[\mu]$ analogous to the
Wess-Zumino-Polyakov action remains to be found out as functional
in the integrating factor $\lam$.

\subsect{Matter field anomalies}

\indent

Having extracted in the previous subsection all the $C$ ghost 
dependence of both $\Xi_{A_n}$ and $\Sigma^{k,i}_{l,A_n}$ it 
remains to study their matter field dependence under the constraints 
\rf{consistC0} and \rf{consistC1} respectively. Due to the
transformation laws of matter fields under diffeomorphisms the ghost 
grading is this time carried by the $\bC$ ghost fields. 
According to the power counting, one finds for $\Xi$,
\be
\Xi_{A_n} = \bC^1\cdots\bC^n\,K_{A_n,\ovl{A}_n}
	\equiv \bC^{\ovl{A}_n}\,K_{A_n,\ovl{A}_n}\ ,
\ee
where $K_{A_n,\ovl{A}_n}$ is a differential polynomial in both 
the $\mu$ and matter fields with $Q_{\Phi\Pi}(K_{A_n,\ovl{A}_n})=0$
and $\dim(K_{A_n,\ovl{A}_n})=2n$. Due to the reality condition 
the cocycle \rf{loco} writes in the matter field sector as
\bea
\Delta_0^{2n+1,\natural}\!& = &\!C^{A_n}
\bC^{\ovl{A}_n}K_{A_n,\ovl{A}_n}\tr(\prt C + \bprt\bC)\nn\\
\!& = &\! c^{A_n}\bc^{\ovl{A}_n}\det(I-\mu\pt\mub)K_{A_n,\ovl{A}_n}
\tr(\prt c + \mu\pt\prt\bc + \bprt\bc + \mub\pt\bprt c).
\lbl{locomat}
\ena
where it is more useful to rewrite the variations with respect to the
true ghos fields. The cocycle condition \rf{nonloc=0}(b) is equivalent to
\be
\delta\lp
c^{A_n}\bc^{\ovl{A}_n}\,\det(I-\mu\pt\mub)\,K_{A_n,\ovl{A}_n}\rp = 0
\ee
and yields
\be
\lp \delta - \tr(\prt c + \bprt\bc)\rp
\,\lp\det(I-\mu\pt\mub)\,K_{A_n,\ovl{A}_n}\rp = c^r \kappa_{r,A_n,\ovl{A}_n} +
\bc^{\ovl{s}}\, \ovl{\kappa}_{\ovl{s},A_n,\ovl{A}_n}. 
\ee
A combination of the derivative operator \rf{derivative} with the
above equation yields
\be
\kappa_{r,A_n,\ovl{A}_n} =
\prt_r\lp\det(I-\mu\pt\mub)\,K_{A_n,\ovl{A}_n}\rp,\cm
\ovl{\kappa}_{\ovl{s},A_n,\ovl{A}_n} = 
\bprt_{\ovl{s}}\lp\det(I-\mu\mub)\,K_{A_n,\ovl{A}_n}\rp,
\ee
which implies
\be
\delta\lp\det(I-\mu\pt\mub)\,K_{A_n,\ovl{A}_n}\rp = 
\prt_r\lp c^r\det(I-\mu\pt\mub)\,K_{A_n,\ovl{A}_n}\rp + 
\bprt_{\ovl{s}}\lp\bc^{\ovl{s}}\det(I-\mu\pt\mub)\,K_{A_n,\ovl{A}_n}\rp.
\lbl{0diffmod}
\ee
The r.h.s. shows that $det(I-\mu\pt\mub)\,K_{A_n,\ovl{A}_n}$ belongs 
to the uncharged scalar density sector of the diff mod $\mbf{d}$ 
cohomology. Given the following variation
\be
\delta\ln\det(I-\mu\pt\mub) = (c\pt\prt
+ \bc\pt\bprt)\ln\det(I-\mu\pt\mub) - \tr(\mu\pt\prt\bc + \mub\pt\bprt c)
\ee
together with eq.\rf{0diffmod} allows to express \rf{locomat} as
\bea
\Delta_0^{2n+1,\natural} \!&=&\!
c^{A_n}\bc^{\ovl{A}_n}\det(I-\mu\pt\mub)K_{A_n,\ovl{A}_n} \tr(\prt c 
+ \bprt\bc)\nn\\
&&\ -\, \delta\lp c^{A_n}\bc^{\ovl{A}_n}\det(I-\mu\pt\mub)K_{A_n,\ovl{A}_n}
\ln\det(I-\mu\pt\mub)\rp.
\lbl{locomatfin}
\ena
Fianlly, we consider the contribution coming from
$\Sigma$. Rewritting the term 
\be 
C^{A_n}\Sigma^{k,i}_{l,A_n}\prt_i \prt_k C^l \equiv
\tr(V\,S),
\ee
inside the cocycle \rf{loco}, the $\delta$-cocycle condition yields
$\delta S = [\prt C,S]$, showing that the $2n-1$ ghost graded quantity
$S$ depends only on the $C$ ghost fields and cannot depend on matter fields.
So no contribution arises from the $\Sigma$ term in the matter sector.

Summing up, the insertion  of the cocycle \rf{locomatfin} 
in the constructive equation
\rf{nat} gives rise to the well-known trace anomaly which breaks the
holomorphic splitting of the partition function when matter fields are
involved.

\indent

\sect{Concluding remarks}

\indent

Even if the physical motivation in studying higher complex dimensional
manifolds is not well stated, the previous considerations can be
regarded as an exercice. The link between Beltrami differentials as
sources of relevant physical tensors in higher dimension is not yet known.
The integrability condition on the $\mu$'s is taken into account and
insures the nilpotency of the BRS operator for the diffeomorphims. 
In fact, the integrability
condition and the nilpotency are actually equivalent. In other words,
demanding an integrable complex structure gives the
nilpotency. Moreover, the integrability condition is preserved under
diffeomorphims. The computation of the generalized Gel'fand-Fuchs
unintegrated $\delta$-cocycles \rf{GF} (modulo $\mbf{d}$) is valid
upon the use of the integrability condition on the Beltrami 
differential because of
the hidden $\mu$ dependence in the $C$ ghost fields and their
diffeomorphism variation $\delta C$. 
However, if one wants to directly check that the anomaly \rf{finam}
(which explicitly depends on $\mu$) is
a $\delta$-cocycle then the use of the integrability condition is
explicitly required.

Once more, we emphasize the role of the holomorphic sector and 
the locality principle which is at the origin of a holomorphically
split diffeomorphism anomalies in the vacuum sector.
The holomorphic property is obtained at the price of restricting to
integrable complex structures.
Contrary to the anomaly given in \cite{LMNS97} for the
so-called ``chiral diffeomorphims", the anomaly computed in the
present work is more general since we have considered the whole 
invariance under reparametrizations. 
However, only the local part of the anomaly has been identified, \ie the
``universal" or ultra-violet contribution independent from the matter fields.
The construction was inspired by the bidimensional
case and relies on the use of 
an integrating factor which is non local in the Beltrami differential.
While in the bidimensional situation, the conformal (Weyl) anomaly 
is equivalent to the well-known holomorphically factorized anomaly
\cite{KLT90}, in higher dimensions, we do not know to which diffeomorphic
invariants the factorized anomalies are equivalent; 
this amounts to finding a globally defined version of the anomalies. 
Moreover, in the presence of matter fields, one recovers the usual trace
anomaly.

Finally, a further careful analysis based on the local 
index theorem of Bismut-Gillet and Soul\'e as pioneered in 
\cite{KLS91a,KLS91b} for the bidimensional case is needed. 
An attempt in that direction has been made in \cite{LMNS97}.

\indent

\noindent
{\bf Acknowledgements :} The authors are grateful to the INFN for some
financial support. One of us (S.L.) wishes to thank the INFN Section of
the University of Genova for its warm hospitality as well as
the Theoretical Physics Group of the University of Amsterdam 
where this work has been intitially carried out under European 
financial support by contract No.ERBCHBICT930301.
Special thanks are due to A.~Blasi for a careful reading of the manuscript.

\indent

\appendix
\renewcommand{\thesection}{Appendix \Alph{section}}
\renewcommand{\theequation}{\Alph{section}.\arabic{equation}}
\renewcommand{\sect}[1]{\setcounter{equation}{0}\section{#1}}

\sect{Diffeomorphism invariance of the integrability condition}\label{invint}

\indent

In this appendix, the diffeomorphism invariance of the integrability
condition on $\mu$ is proved. With $\mbf{\mu}^{\varphi}$ defined by
eq.\rf{mufi} of the main text, let us compute directly the integrability 
condition and similarly to the first order differential operator
$\mbf{\cal L}$ defined by eq.\rf{L}, one set 
$\mbf{\cal L}^{\varphi}= \fbprt - \mbf{\mu}^{\varphi}$. The (0,2)-form
with (1,0)-vector values reads
\bea
\fbprt\mbf{\mu}^{\varphi} - (\mbf{\mu}^{\varphi})^2 &=& 
\mbf{\cal L}^{\varphi} \mbf{\mu}^{\varphi}
\ =\ - \blp \mbf{\cal L}^{\varphi}\ovl{\varphi} \brp \pt
\blp \mbf{\cal L}^{\varphi}(\mu\circ\varphi) \brp
\pt \blp \prt\varphi +\prt\ovl{\varphi}\pt(\mu\circ\varphi)\brp^{-1}\pt\prt
\nn\\
&=& - \blp \mbf{\cal L}^{\varphi}\ovl{\varphi} \brp \pt
\blp ( \mbf{\cal L}^{\varphi}\varphi )\pt \prt\, \mu
+ (\mbf{\cal L}^{\varphi}\ovl{\varphi})\pt \bprt\, \mu \brp
\pt \blp \prt\varphi +\prt\ovl{\varphi}\pt(\mu\circ\varphi)\brp^{-1}\pt\prt
\nn\\
&=& - \blp \mbf{\cal L}^{\varphi}\ovl{\varphi} \brp \pt
\blp (\mbf{\cal L}^{\varphi}\ovl{\varphi}) \pt {\cal L} \mu \brp
\pt \blp \prt\varphi+\prt\ovl{\varphi}\pt(\mu\circ\varphi)\brp^{-1}\pt\prt\ ,
\lbl{last}
\ena
where the third equality is obtained by using 
$\prt = \prt\varphi\pt\prt + \prt\ovl{\varphi}\pt\bprt$ and
$\fbprt = \fbprt \varphi\pt \prt + \fbprt \ovl{\varphi}\pt\bprt$,
while the fourth one comes from the identity 
$ \mbf{\cal L}^{\varphi} \varphi = - (\mbf{\cal L}^{\varphi} \ovl{\varphi})
\pt(\mu\circ\varphi)$.
Now, the integrability condition for $\mu$ writes
\be
\fbprt\mbf{\mu} - \mbf{\mu}^2\ =\ 0 \Llra {\cal L}_\ib\, \mu_\jb^k \ =\
{\cal L}_\jb\, \mu_\ib^k\ ,
\ee 
and shows that ${\cal L}\mu$ is a mixed tensor, symmetric in the lower
indices. This symmetry implies the vanishing of the last expression
\rf{last},
\be
\fbprt\mbf{\mu} - \mbf{\mu}^2\ =\ 0 \Lra
\fbprt\mbf{\mu}^{\varphi} - (\mbf{\mu}^{\varphi})^2\ =\ 0 \ .
\ee

\sect{On the equivalence between complex structures}\label{equivstruct}

\indent

Any diffeomorphism $\varphi$ when expressed in terms of its local 
representative will induce a smooth change of local coordinates. This change
of variables can be written in matrix notation (with matrix entries),
as well the inverse change of variables, as
\be
\lp \begin{array}{c}
\prt \varphi \\[1mm] \hline \\[-5mm] \bprt\varphi \end{array} \right|
\left. \hs{-1.2}\begin{array}{c}
\prt \ovl{\varphi} \\[1mm] \hline \\[-5mm] \bprt\ovl{\varphi}
\end{array} \rp\ , \cm \cm
\lp \begin{array}{c}
\mbox{A} \\[1mm] \hline \\[-5mm] \mbox{C} \end{array} \right|
\left. \hs{-1.2}\begin{array}{c}
\mbox{B} \\[1mm] \hline \\[-5mm] \mbox{D}
\end{array} \rp\ ,
\ee 
with the following relations
\bea
&&\mbox{A}\pt\prt \varphi + \mbox{B}\pt\bprt\varphi\ =\ I_w\ , \cm\cm
\prt \varphi\pt\mbox{A} + \prt \ovl{\varphi}\pt\mbox{C}\ =\ I_z\ , \nn\\
&&\mbox{A}\pt\prt\ovl{\varphi} + \mbox{B}\pt\bprt \ovl{\varphi}\
=\ 0\ \ , \cm\cm\,
\prt \varphi\pt\mbox{B} + \prt \ovl{\varphi}\pt\mbox{D}\ =\ 0\ \ ,\nn\\[-4mm]
&&{}\lbl{invmat}\\[-3mm]
&&\mbox{C}\pt\prt \varphi + \mbox{D}\pt\bprt\varphi\ =\ 0\ \ , \cm\cm\,
\bprt\varphi\pt\mbox{A} + \bprt \ovl{\varphi}\pt\mbox{C}\ =\ 0\ \ , \nn\\
&&\mbox{C}\pt\prt\ovl{\varphi} + \mbox{D}\pt\bprt\ovl{\varphi}\ =\ 
I_{\ovl{w}}\ , \cm\cm
\bprt\varphi\pt\mbox{B} + \bprt \ovl{\varphi}\pt\mbox{D}\ =\ I_{\bz}\ .\nn
\ena
Starting with the equation in the target chart with respect to
$\varphi$, $(w,\ovl{w})=\varphi\zbz$,
\be
\blp \prt_{\ovl{w}} - (\mu\circ\varphi)\pt \prt_w \brp Z\ =\ 0\ ,
\lbl{targetbeltra}
\ee
and by using the two last identities of the right culumn of
the above equations \rf{invmat}, one gets in the source chart of $\varphi$,
\be
\blp \prt_{\bz} - (\bprt\varphi + \bprt\ovl{\varphi}\pt(\mu\circ\varphi))
\pt(\mbox{B}\pt\prt_{\bz}+\mbox{A}\pt\paZ)
\brp (Z\circ\varphi)\ =\ 0 \ ,
\ee
and by eq.\rf{finite} of the main text combined with the two first
identities of the right culumn of \rf{invmat}, one has
\be
\blp \prt_{\bz} - (\mu^{\varphi})\pt\paZ 
+ (\mu^{\varphi})\pt\prt\ovl{\varphi}\pt
\biggl[ (\mbox{C}\pt\paZ + \mbox{D}\pt\prt_{\bz}) 
- (\mu\circ\varphi)\pt (\mbox{A}\pt\paZ +
\mbox{B}\pt\prt_{\bz}) \biggr] \brp (Z\circ\varphi) \ =\ 0.
\ee
The terms involving $\prt\ovl{\varphi}$ turnos out to be proportional to
eq.\rf{targetbeltra} due to the laws for the change of variables, and
thus the last parenthesis vanishes. It remains
\be
\blp \prt_{\bz} - (\mu^{\varphi})\pt\paZ \brp (Z\circ\varphi) \ =\ 0,
\ee 
which shows that $(Z\circ\varphi)$ also provide $n$ linearly independent
solutions of the Beltrami equations on the source of the
diffeomorphism $\varphi$, see eq.\rf{sourcebeltra} in the text which is
here recalled
\be
\blp \prt_{\bz} - (\mu^{\varphi})\pt\paZ \brp Z^{\varphi}\ =\ 0\ .
\ee
Thus the mapping $(Z^{\varphi},\bZ^{\varphi})\lra\ZBZ$ is indeed
bi-holomorphic, see Theorem 5.3 in \cite{Kod86}. In other words,
this result states the equivalence between the complex structures 
parametrized by $\mbf{\mu}$ and $\mbf{\mu}^{\varphi}$.

\sect{True holomorphic divergence}\label{divfree}

\indent

In the following the computation of the divergence \rf{ZWid} is
performed. In doing so, several results are required and are listed
hereafter.

Firs, the computation of the inverse matrix
\be
\lp \begin{array}{c}
I \\[1mm] \hline \\[-5mm] \mu \end{array} \right|
\left. \hs{-1.2}\begin{array}{c}
\mub \\[1mm] \hline \\[-5mm] I \end{array} \rp^{-1}
=  \lp \begin{array}{c}
(I - \mub\pt\mu)^{-1}\\[1mm] \hline \\[-5mm] 
-(I-\mu\pt\mub)^{-1}\pt\mu \end{array} \right|
\left. \hs{-1.2}\begin{array}{c}
-(I - \mub\pt\mu)^{-1}\pt\mub \\[1mm] \hline \\[-5mm] 
(I-\mu\pt\mub)^{-1} \end{array} \rp\ ,
\ee
yields in particular the following two identities
\be
(I - \mub\pt\mu)^{-1} - \mub\pt(I-\mu\pt\mub)^{-1}\pt\mu \equiv I,\cm
(I - \mub\pt\mu)^{-1}\pt\mub - \mub\pt(I-\mu\pt\mub)^{-1}\equiv 0\ .
\lbl{invid}
\ee
Second, one will use the identities
\be
\prt_{Z^k}\lp\frac{1}{\cal J} \lam^k_i \rp + 
\prt_{Z^\lb}\lp\frac{1}{\cal J} \mub_i^\rb\ \lamb_\rb^{\,\lb} 
\rp \equiv 0\ ,\cm
\prt_{Z^k}\lp\frac{1}{\cal J} \mu_\ib^r\lam^k_r \rp + 
\prt_{Z^\lb}\lp\frac{1}{\cal J} \lamb_\ib^{\,\lb} \rp \equiv 0\ ,
\lbl{Jacobimulti}
\ee
coming from the Jacobi multipliers.

We are now in the position for computing the divergences \rf{ZWid} of the
main text. Taking into account \rf{compa} and \rf{nabla}
one can show,
\be
\frac{(I-\mu\pt\mub)_\lb^\jb\,\lamb_\jb^\rb }{\det(\lamb(I - \mu\pt\mub))}
\prt_{Z^\rb} \lp \frac{1}{\det\lam}(\lam^{-1})_k^s\Theta_s^\lb \rp
=
\frac{1}{\cal J} (\lam^{-1})_k^s \lp \cL_\ib \Theta^\ib_s -
\prt_r\mu^r_\ib\Theta^\ib_s - \prt_s\mu_\ib^r\Theta^\ib_r \rp
\ee
which relates to the Ward id's \rf{Wid} by virtue of the
non-singularity of $\lam$. 
It is straightforward to show that
\be
\prt_{Z^\rb} \lp \frac{1}{\det(\lamb(I - \mu\pt\mub))} 
(I - \mu\pt\mub)\pt\lam^\rb \rp \equiv 0\ ,
\ee
thanks to the identities \rf{Jacobimulti} and \rf{invid}.

\sect{The $\hdelta$ Cohomology}\label{coho}

\indent

The $\hdelta$ operator, given by formulae \rf{hdel} in the text,
is defined on the space of local functions considered, according to the
power counting, as differential polynomials 
in both the matter and the $\Phi$-$\Pi$ charged fields, and as analytic
in the components of the Beltrami differential which are of zero dimension. 
The following set of fields
$$\chi\equiv f_0 \cup f_1 \cup \beta \cup \{\lam,\ovl{\lam}\},\cm
f_0\equiv\{\phi,\mu,\mub\},\cm f_1\equiv\{c,\bc\},\cm
\beta\equiv\{\gamma,\eta,\ovl{\eta},\zeta,\ovl{\zeta}\},$$
will serve as generators for our space of local functions.
Thanks to the definition of $\hdelta$, these local functions do
not contain any underived ghosts $c$'s.

Let us recall that the fields and their derivatives will be considered 
as independent coordinates, and in practice, will play the role of
creation operators, while the annihilators will be the formal
derivatives with respect to these coordinates, both acting 
on the Fock space structure of the space of local functions.
This Fock space is graded according to the ghost grading.
For any operator, its adjoint will be given by 
the formal replacement of the derivative with respect to some
coordinate by the formal multiplication with respect the same
coordinate and vice versa.  

According to this rule, let us now introduce the following 
self-adjoint operator,
\be
\nu\ =\ \sum_{|I|,|\ovl{J}|\geq 0} (|I|+|\ovl{J}|) \lp
\sum_{f_1} <\prt_I\prt_{\ovl{J}}\,f_1\zbz
\DER{\prt_I\prt_{\ovl{J}}\,f_1} \rp
\lbl{nu}
\ee
whose eigenvalues will provide the order of the derivatives of the
ghost fields. The operator $\nu$  will decompose the space of local
functions into a direct sum of subspaces according to its eigenvalues
while the operator $\hdelta$ will be filtered according to
\be
\grt{[}\nu , \hdelta \grt{]}\ =\ \sum_{|I|,|\ovl{J}|\geq 0} (|I|+|\ovl{J}|)\,
\hdelta(|I|+|\ovl{J}|)\ ,\cm
\hdelta \ =\ \sum_{|I|,|\ovl{J}|\geq 0}\hdelta(|I|+|\ovl{J}|)\ .
\lbl{filtra}
\ee
The general spectral method \cite{Dix76,Ba88}
insures that the $\hdelta$-cohomology 
is isomorphic to the solutions $\tld{\Delta}$ of the system 
\be
\left\{ \begin{array}{c}
\hdelta(|I|+|\ovl{J}|) \tld{\Delta} \zbz=0 \\[2mm]
\hdelta^{\dagger}(|I|+|\ovl{J}|) \tld{\Delta} \zbz=0
\end{array} \right. ,\cm  |I|+|\ovl{J}|\geq 0.
\lbl{iso}
\ee
In other words, $\tld{\Delta}$ is in the kernel of the Laplacians
$\left\{ \hdelta(|I|+|\ovl{J}|),\hdelta^{\dagger}(|I|+|\ovl{J}|)\right\}$.

In the following, we present in full details the solution of the
cohomology by means of the spectral sequences.
The zero eigenvalue of $\nu$ will select the part of $\hdelta$ in
which no ghost field is involved, namely
\bea
\hdelta(0) &=& \sum_{|I|,|\ovl{J}|\geq 0} \lp
\sum_{\beta_{-1}} 
<\prt_I\prt_{\ovl{J}} \frac{\delta\Gamma^{\mbox{\tiny Cl}}_0}{\delta f_0\zbz},
\DER{\prt_I\prt_{\ovl{J}}\,\beta_{-1}}> \right.\nn\\
&&\left.+ \sum_{\beta_{-2}}
<\prt_I\prt_{\ovl{J}} X(f_0,\beta_1)\zbz,
\DER{\prt_I\prt_{\ovl{J}}\,\beta_{-2}}>\rp\ .\lbl{delta0}
\ena
Note that the latter is a nilpotent operator and is an annihilator
with respect to the external sources coupled to the BRS-variations.
One will prove that the $\hdelta$-cohomology does not actually 
depend on the external sources coupled to the BRS-variations
by performing a double-filtration. 
One first filtrates according to the counting operator 
\be
{\cal N}(\beta)\ =\ \sum_{|I|,|\ovl{J}|\geq 0}
<\prt_I\prt_{\ovl{J}}\beta\zbz,\DER{\prt_I\prt_{\ovl{J}}\beta}>,
\ee
with respect to only {\em one} of the external sources 
$\beta\equiv\{\gamma,\eta_k^\ib,\ovl{\eta}_\jb^l,\zeta_i,\ovl{\zeta}_\ib\}$. 
Next, the second step concists in a filtration with respect to the
full counting operator
\be
{\cal N}\ = \sum_{\chi} \sum_{|I|,|\ovl{J}|\geq 0} 
<\prt_I\prt_{\ovl{J}}\,\chi\zbz,\DER{\prt_I\prt_{\ovl{J}}\chi}> .
\ee
By virtue of the isomorphism \rf{iso}, the Laplacian corresponding 
to the given operator in the null subsector of the double filtration 
$\left[{\cal N},\left[{\cal N}(\beta),\hdelta(0)\right]\right]$, 
shows that the $\hdelta$-cohomology does not depend on any negatively
ghost graded external sources $\beta$, thus restricts the set of
generators to $\chi\equiv f\cup\{\lam,\ovl{\lam}\}$, $f\equiv f_0\cup f_1$.

The second operator $\hdelta(1)$ in the filtration \rf{filtra} 
will necessarily contain the ghost part, $\hdelta_{\tiny Gh}$
of $\hdelta$ since $[\nu,\hdelta_{\tiny Gh}]=\hdelta_{\tiny Gh}$. 
The latter will play a central role in the 
computation of the cohomology, For both this reason and 
the sake of compacteness in the notation it is written as
\bea
\hdelta_{\tiny Gh}&=& 
\sum_{|I|\geq 0}\delta^{M+N}_I \frac{(I+1_\al)!}{(M+1_\al)!N!}\ 
(\prt_{M+1_\al}\, c^{\beta} \prt_{N+1_{\beta}}\,c^{\rho})\zbz 
\DER{\prt_{I+1_\al}\, c^{\rho}},\\
&&\al,\beta,\rho=1,\dots,n,\dots,2n,\cm 
\al \equiv \left\{\begin{array}{l}
i,\cm \mbox{\nms for}\ \al = 1,\dots,n\ ,\\
\ib,\cm \mbox{\nms for}\ \al = n+1,\dots,2n ;\end{array}\right.\nn\\
&& I=(i_1,\dots,i_{2n}),\cm i_\al = \left\{\begin{array}{l}
i_k,\cm \mbox{\nms for}\ \al = k = 1,\dots,n\ ;\\
\ib_l,\cm \mbox{\nms for}\ \al = n+1,\dots,2n,\cm l=\al-n
\end{array}\right.\nn\\
&& c^\al\ =\ \left\{\begin{array}{l}
c^i,\cm \mbox{\nms for}\ \al = 1,\dots,n\ ;\\
\bc^{\,\ib},\cm \mbox{\nms for}\ \al =
n+1,\dots,2n.\end{array}\right.\nn
\ena
One can now write
\be 
\hdelta(1) \equiv \prt_\al\, c^{\beta}\zbz U^\al_{\beta} + R(1)
\lbl{filtra1}
\ee
where, according to the ghost grading, one can make the following 
decomposition for the operator $U^\al_{\beta} \equiv (U_0)^\al_{\beta}
+ (U_1)^\al_{\beta}$, where $(U_0)^\al_{\beta}$ is given by the
Lie derivatives of the fields described as components of geometrical 
objects and $(U_1)^\al_{\beta}$ comes from $\hdelta_{\tiny Gh}$. 
For instance, in the case of matter fields 
$U^i_j\equiv S^i_j$ one has
\bea
(S_0)_j^i &=&\! \sum_{|M|,|\ovl{N}|\geq 0} 
\lp (m_i +1) \sum_{\chi\equiv f_0\cup\{\lam,\ovl{\lam}\}}
<\prt_{M+1_j}\prt_{\ovl{N}}\,\chi\zbz,\DER{\prt_{M+1_i}\prt_{\ovl{N}}\,\chi}>
\right.\nn\\
&&\hs{10}+\ \sum_{A_{n-1}} \delta_{A_{n-1}}^{A_{r+1_j}}
<\prt_M\prt_{\ovl{N}}\,\phi_{A_{r+1_j}}\zbz,
\DER{\prt_M\prt_{\ovl{N}}\,\phi_{A_{r+1_i}}}>\nn\\
&&\hs{-10}+\ 
 \prt_M\prt_{\ovl{N}}\,\lam^r_j\zbz\DER{\prt_M\prt_{\ovl{N}}\lam^r_i}
+ \prt_M\prt_{\ovl{N}}\,\mub^{\ovl{s}}_j\zbz
\DER{\prt_M\prt_{\ovl{N}}\,\mub{\ovl{s}}_i}\\
&&\left.\hs{10} - \ \prt_M\prt_{\ovl{N}}\,\mu^i_{\ovl{s}}\zbz
\DER{\prt_M\prt_{\ovl{N}}\,\mu^j_{\ovl{s}}}\rp\ ;\nn\\
(S_1)_j^i &=& \! \sum_{|M|\geq 0} 
 (m_i +1) \prt_{M+1_j}\,c^\al\zbz \DER{\prt_{M+1_i}\,c^\al}\nn\\
&&\hs{40}-\ \sum_{|M|\geq 1}\prt_M\,c^i\zbz \DER{\prt_M\,c^j}\ .
\ena
Note in addition that the trace $S_i^i=\tr S$ is nothing else but the 
``little z" indices counting operator
\be
\tr S\ =\ N_{\prt} + N_{\lambda} + N_{\mub} + N_{\phi} - N_{\mu} - N_{c} \, 
\equiv N_{z}(\downarrow)-N_{z}(\uparrow)
\lbl{count}
\ee
One has ${S^{\dagger}}_i^j={S^{\sim}}_i^j\equiv S_j^i$ and 
$\ovl{S}\equiv N_{\bz}(\downarrow)-N_{\bz}(\uparrow)$ with 
$\bar{S}^{\dagger}=\bar{S}^{\sim}$.
In full generality, one has for the ghost contribution
\be
(U_1)^\al_{\beta} = \sum_{|I|\geq 0}\lp (i_\al+1)\,\prt_{I+1_{\beta}}
\,c^{\rho}\zbz 
\DER{\prt_{I+1_\al}\,c^{\rho}} - \prt_{I+1_{\rho}+1_{\sigma}}\,c^\al
\DER{\prt_{I+1_{\rho}+1_{\sigma}}\,c^{\beta}}\rp ,
\ee
and the remaining piece $R(1)$ concerns only the ghost derivatives of
order greater or equal to three, 
\bea
\leqa{ R(1)\ = \sum_{|M|\geq 0} \prt_{M+1_\al+1_{\beta}} c^{\rho}\zbz\,
R_{\rho}^{M+1_\al+1_{\beta}}\zbz\ , }\\
R_{\rho}^{M+1_\al+1_{\beta}}\zbz\! &=&\!\! \sum_{|N|\geq 0}
\mbox{$\frac{(M+N+1_\al+1_{\beta}+1_{\gamma})!}{(M+1_\al+1_{\beta})!(N+1_{\gamma})!}$}\
\prt_{N+1_{\rho}+1_{\gamma}} c^{\sigma}\zbz
\DER{\prt_{M+N+1_\al+1_{\beta}+1_{\gamma}}c^{\sigma}}\ .\nn
\ena
Furthermore, the nilpotency, $\hdelta(1)^2=0$, allows one to
apply once more the spectral sequence analysis to the operator $\hdelta(1)$.
The filtration will be performed with respect to the following
counting operator
\be
\nu'\ =\ 1 + \prt_\al\,c^{\beta}\zbz \DER{\prt_\al\,c^{\beta}}\ .
\lbl{nu'}
\ee
Remarkably, the cohomology reduces to the following finite filtration
\be
\grt{[} \nu' , \hdelta(1) \grt{]}\ =\ \sum_{a=1}^2 a \, \hdelta'(a)\ ,
\cm\mbox{\nms where}\ \left\{\begin{array}{l}
\hdelta'(1)\ =\ R(1),\\
\hdelta'(2)\ =\ \prt_\al\, c^{\beta}\zbz U^\al_{\beta}\ ,
\end{array}\right.
\lbl{filtra'}
\ee
and $\hdelta(1) = \hdelta'(1) + \hdelta'(2)$.
According to \rf{iso}, one has to solve the system
\be \left\{ \begin{array}{l}
\hdelta'(a) \tld{\Delta} \zbz\ =\ 0 \\
\hdelta'^{\dagger}(a) \tld{\Delta} \zbz\ =\ 0
\end{array} \right.\ ,\cm \mbox{\nms for}\ a=1,2,
\lbl{iso'}
\ee
which turns out to be equivalent to
\be
<\! \tld{\Delta}\zbz| \grt{\{} \hdelta'^{\dagger}(a) , \hdelta'(a) \grt{\}}
|\tld{\Delta}\zbz\!> = || \hdelta(a) \tld{\Delta}\zbz||^2 
+ || \hdelta'^{\dagger}(a) \tld{\Delta}\zbz||^2\stackrel{!}{=} 0 \ ,
\lbl{lapla}
\ee
where the scalar product is the one induced by the
definition of the adjoint operator in the space of local functions.
For $a=1$, the lowest order of formal derivatives with respect to
the ghost derivatives in the Laplacian involved in eq.\rf{lapla} 
can be seen of order three.
So, the cocycle $\tld{\Delta}$ depends on the first and
second order derivatives of the ghost fields.
The second filtration $a=2$, selects the scalar sector, and
eliminates the second order ghost derivatives.
It is then useful to perform the following change
of variables in the first order derivatives of the ghost
fields. Define, in matrix notation, the $n^2$ variables of ghost grading
one,
\be
\tld{\Omega}\zbz\ =\ (\prt c + \prt \bc\pt\mu) \zbz =
(\hdelta\lam\pt\lam^{-1})\zbz ,
\ee
with $(\tr S)\tld{\Omega}=0$, and note the $\hdelta$-coboundary
\be
\tr\tld{\Omega}\zbz\ =\ \hdelta\ln\det\lam\zbz .
\ee
One has $\hdelta\tld{\Omega}\zbz = \hdelta'(2)\,\tld{\Omega}\zbz
= \tld{\Omega}^2\zbz$, and
\be
\hdelta\tld{\Omega}^{2k+1}\zbz = 
\tld{\Omega}^{2k+2}\zbz,
\ee
showing that the $\tld{\Omega}^{2k+2}$ are $\hdelta$-coboundaries. 
Taking the trace
in order to project onto the scalar sector, one readily checks that 
\be
\hdelta\tr\tld{\Omega}^{2k+1}\zbz\ =\ \tr\tld{\Omega}^{2k+2}\zbz\ =\
-\tr\tld{\Omega}^{2k+2}\zbz\ =\ 0\ .
\ee
This yields the following theorem which is the main result of this appendix.

\vskip .5cm\noindent
{\bf Theorem.} {\em In the scalar sector of the 
space of analytic functions in the fields and their derivatives, 
the ghost sectors with grading $p$
of the ${\hdelta}$-cohomology are generated by the following 
non-trivial cocycles}
\bea
&& \begin{array}{ll}
\mbox{\nms for even}\ p\ :\ &
\tr\tld{\Omega}^{2r+1}\zbz\,\tr\ovl{\tld{\Omega}}{}^{2s+1}\zbz\ ,\\[2mm]
\mbox{\nms for odd}\ p,\ :\ &
\tr\tld{\Omega}^{2k+1}\zbz,\\
\end{array} \nn
\ena
{\em while the cocycle}
$\tr\tld{\Omega}$ {\em can be reabsorbed by completing the set of
generators with $\ln\det\lam$ seen as an independent 
variable. The same statements hold true for the complex conjugate expressions.}


\newpage

\begin{thebibliography}{10}
\bibliographystyle{unsrt}

\bibitem{LMNS97}
A.~Losev, G.~Moore, N.~Nekrasov, and S.~Shatashvili.
\newblock Chiral {L}agrangians, anomalies, supersymmetry and holomorphy, {\tt
  hep-th/9606082}.
\newblock {\em Nucl. Phys. {\bf B484}}, page 196, 1997.

\bibitem{BB87}
L.~Baulieu and M.~Bellon.
\newblock Beltrami parametrization in string theory.
\newblock {\em Phys. Lett. {\bf B196}}, page 142, 1987.

\bibitem{Bec88}
C.~Becchi.
\newblock On the covariant quantization of the free string: the conformal
  structure.
\newblock {\em Nucl. Phys. {\bf B304}}, page 513, 1988.

\bibitem{KLT90}
M.~Knecht, S.~Lazzarini, and F.~Thuillier.
\newblock Shifting the {W}eyl anomaly to the chirally split diffemorphism
  anomaly in two dimensions.
\newblock {\em Phys. Lett. {\bf B251}}, pages 279--283, 1990.

\bibitem{KLS91a}
M.~Knecht, S.~Lazzarini, and R.~Stora.
\newblock On holomorphic factorization for free conformal fields.
\newblock {\em Phys. Lett. {\bf B262}}, pages 25--31, 1991.

\bibitem{KLS91b}
M.~Knecht, S.~Lazzarini, and R.~Stora.
\newblock On holomorphic factorization for free conformal fields {II}.
\newblock {\em Phys. Lett. {\bf B273}}, pages 63--66, 1991.

\bibitem{Vaf94}
C.~Vafa.
\newblock {\em Mirror transform and string theory}.
\newblock Talk given in the {\em Geometry and Topology Conference}, April '93,
  Havard in honour of R. Bott, hep-th/940315, March 1994.

\bibitem{BCOV94}
M.~Bershadsky, S.~Cecotti, H.~Ooguri, and C.~Vafa.
\newblock Kodaira-spencer theory of gravity and exact results for quantum
  string amplitudes.
\newblock {\em Comm. Math. Phys. {\bf 165}}, pages 311--428, 1994.

\bibitem{BS94}
M.~Bershadsky and V.~Sadov.
\newblock {\em Theory of {K}{\"a}hler gravity}.
\newblock hep-th/9410011, 1994.

\bibitem{LM94}
J.M.F. Labastida and M.~Marino~(Santiago de~Compostela~U.).
\newblock Type {B} topological matter, {K}odaira-{S}pencer theory, and mirror
  symmetry.
\newblock {\em Phys. Lett. {\bf B333}}, pages 386--395, 1994.

\bibitem{Zuc97}
R.~Zucchini.
\newblock {\em The quaternionic structure of 4d conformal field theory}.
\newblock {\tt gr-qc/9707048}, July 1997.

\bibitem{Ba88}
G.~Bandelloni.
\newblock Diffeomorphism cohomology in quantum field theory models.
\newblock {\em Phys. Rev. {\bf D38}}, page 1156, 1988.

\bibitem{BaLa93}
G.~Bandelloni and S.~Lazzarini.
\newblock Diffeomorphism cohomology in {B}eltrami parametrization.
\newblock {\em Jour. Math. Phys. {\bf 34}}, page 5413, 1993.

\bibitem{BaLa95}
G.~Bandelloni and S.~Lazzarini.
\newblock Diffeomorphism cohomology in {B}eltrami parametrization {II}: the
  1-forms.
\newblock {\em Jour. Math. Phys. {\bf 36}}, pages 1--29, 1995.

\bibitem{Kod86}
K.~Kodaira.
\newblock {\em Complex {M}anifolds and {D}eformation of {C}omplex
  {S}tructures}.
\newblock Comprehensive Studies in Mathematics. Spinger-Verlag, New-York, 1986.

\bibitem{Sto88}
R.~Stora.
\newblock The role of locality in string theory.
\newblock In G.~'t~Hooft and al., editors, {\em Non perturbative quantum field
  theory}, NATO ASI Ser.B, Vol.185. Plenum Press, 1988.

\bibitem{Laz90}
S.~Lazzarini.
\newblock {\em Sur les mod\`eles conformes lagrangiens bidimensionnels}.
\newblock PhD thesis, Universit\'e de Savoie, April 1990.
\newblock unpublished.

\bibitem{Kos60}
J.-L. Koszul.
\newblock {\em Lectures on fibre bundles and differential geometry}.
\newblock Reissued 1965. Tata Institute of Fundamental Research, Bombay, 1960.

\bibitem{PR81}
O.~Piguet and A.~Rouet.
\newblock Symmetries in perturbative quantum field theory.
\newblock {\em Phys. Rep. {\bf 76}}, pages 1--77, 1981.

\bibitem{Dix76}
J.~Dixon.
\newblock {\em Cohomology and Renormalization of Gauge Theories I-II-III}.
\newblock unpublished, 1976-1979.

\bibitem{Fucks86}
D.B. Fucks.
\newblock {\em Cohomology of infinite Dimensional Algebra}.
\newblock Consultant Bureau, New York, 1986.

\end{thebibliography}
\end{document}